\documentclass[amsmath,amssymb,english,aps,pra, twocolumn,floatfix]{revtex4-1}
\usepackage[T1]{fontenc}
\usepackage[latin9]{inputenc}
\usepackage{amsmath}
\usepackage{amssymb}
\usepackage{graphicx}
\usepackage{babel}
\usepackage{color}
\usepackage{braket}
\usepackage{subfigure}
\usepackage[normalem]{ulem}
\usepackage[colorlinks,bookmarks=false,citecolor=magenta,linkcolor=blue,urlcolor=black]{hyperref}
\usepackage{textcomp}

\usepackage{bm}
\usepackage{physics}

\usepackage[colorinlistoftodos,prependcaption,textsize=small]{todonotes}

\begin{document}

\title{Ergodicity breaking with long range cavity induced quasiperiodic interactions}
\author{Piotr Kubala}
\email{piotr.kubala@doctoral.uj.edu.pl}
\affiliation{Institute of Theoretical Physics, Jagiellonian University, \L{}ojasiewicza 11, 30-348 Krak\'ow, Poland }
\author{Piotr Sierant}
\affiliation{The  Abdus  Salam  International  Center  for  Theoretical  Physics, Strada  Costiera  11,  34151,  Trieste,  Italy}
\affiliation{Institute of Theoretical Physics, Jagiellonian University, \L{}ojasiewicza 11, 30-348 Krak\'ow, Poland }
\author{Giovanna Morigi}
\affiliation{Theoretical Physics, Saarland University, Campus E2.6, D--66123 Saarbr\"ucken, Germany}
\author{Jakub Zakrzewski}
\affiliation{Institute of Theoretical Physics, Jagiellonian University, \L{}ojasiewicza 11, 30-348 Krak\'ow, Poland }
\affiliation{Mark Kac Complex
Systems Research Center, Jagiellonian University, \L{}ojasiewicza 11, 30-348 Krak\'ow,
Poland. }
\email{jakub.zakrzewski@uj.edu.pl}

\begin{abstract}
Many-body localization (MBL) behavior is analyzed {in an extended Bose-Hubbard model with quasiperiodic infinite-range interactions. No additional disorder is present.}
{Examining level statistics and entanglement entropy of eigenstates we show that a significant fraction of eigenstates of the system is localized in the presence of strong interactions. In spite of this, our results suggest that the system becomes ergodic in the standard thermodynamic limit in which the energy of the system is extensive. At the same time, the MBL regime seems to be stable if one allows for a super-extensive scaling of the energy.   }
{We show that our findings can be experimentally verified by studies of time dynamics in many-body cavity quantum electrodynamics setups. The ``quench spectroscopy'' is a particularly effective tool that allows us to systematically study energy dependence of time dynamics and to investigate a mobility edge in our system. }
\end{abstract}

\maketitle

\section{Introduction}

Many-body localization (MBL) is a robust way of  ergodicity breaking in interacting many-body systems. The seminal
early works \cite{Gornyi05,Basko06, Oganesyan07}
show that such a behavior is associated with the presence 
of strong disorder in the system. Hundreds of studies in the last 15 years addressed 
various aspects of MBL so we  refer the reader to reviews of the 
problem \cite{Nandkishore15, Luitz17b, Alet18,Abanin19, Gopalakrishnan20} 
Despite the numerous works, even the issue of the existence of MBL 
in thermodynamic limit is not settled yet.
On one side there {is a proof of existence MBL} in certain one-dimensional (1D) spin systems \cite{Imbrie16,Imbrie16a}
under assumptions regarding properties of energy levels. On
the other side recent numerical studies
put into question \cite{Suntajs19} 
the very existence of MBL as a stable dynamical phase of matter or point out difficulties in its unambiguous 
verification  \cite{Panda20,Sierant20b}.
{Subsequent works provided further arguments against the stability of MBL phase \cite{Kiefer20, Sels20},
or supported the scenario of stable MBL phase \cite{Luitz20} indicating that earlier {predictions} \cite{Luitz15} of critical disorder
strengths may {underestimate the real value} \cite{Sierant20p} (see also \cite{Doggen18, Chanda20t}). 
Those works concentrate mainly on the model of
spin$-1/2$ systems with short range interactions and a uniform random disorder present. 

Instead, the experimental attempts 
to observe signatures of MBL often employ a quasiperiodic  potential \cite{Choi16,Luschen17,Luschen18}
realized by superimposing two incommensurate optical lattices, the  primary one holding the atoms (allowing for
the tight binding approximate description)  and the secondary one producing the disorder 
by weakly perturbing local minima  of
the primary lattice. The quasiperiodic potential leads to MBL similarly to the uniform random disorder.
It is, however, claimed that the universality class of 
the transition from delocalized, extended phase to MBL phase  is dependent on the disorder type
\cite{Khemani17, Maksymov20}. Similarly,
the range of interactions plays a role as well as the dimension of the problem. We 
shall restrict ourselves to 1D case, still here, the existence of MBL seem to be dependent on the interaction
range. In particular for power-law decaying interactions $V\propto 1/r^\alpha$ where $r$ is the distance between 
interacting particles it is claimed that MBL exists for $\alpha>1$ \cite{Burin15,Burin15b}. 
This condition is naturally broken in a system with 
 cavity mediated all-to-all interactions \cite{Sierant19c}. The results presented by two of us in this work, while valid for a finite 
system may nevertheless be fragile in the thermodynamic limit as pointed out by 
\cite{MaksymovBurin20}.

The aim of this work is to investigate ergodicity
breaking in the presence of all-to-all
cavity mediated interactions in the quasi-periodic disorder resulting from the mismatch between
the cavity mode wavelength and the wavelength of the optical lattice parallel to the cavity axis.
Hence, contrary to \cite{Sierant19c} we do not assume any additional random or quasi-periodic on-site
disorder. We shall show that while an MBL regime may be seen for a finite size system, the
thermodynamic limit is subtle and the results depend on the details of the model implementation.
The paper is organized as follows. The model and its Hamiltonian are discussed in the next Section
followed by {the presentation of} spectral properties of the system. Then we discuss the time dynamics and {consider different}  
possible quenches in the system. As we shall see some care must be taken in order to excite the
system in the controlled way.

\section{Hamiltonian}

\begin{figure}[htbp]
    \centering
    \includegraphics[width=\linewidth]{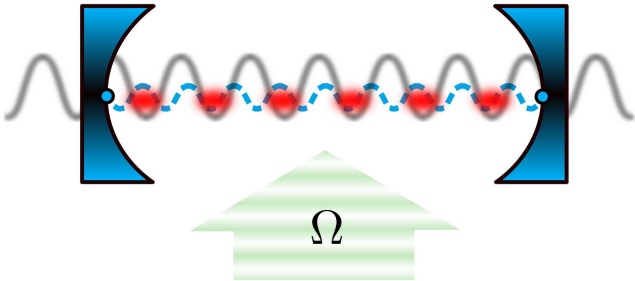}
    \caption{Quasi-periodic infinite-range interactions can be realised by a quantum gas of bosons in an optical lattice  (solid, grey line) and strongly coupling to a standing-wave mode of a high-finesse optical resonator (dashed blue line). The dipolar transitions couple dispersively with the cavity mode and with a transverse laser beam. As a result the multiply scattered photons effectively mediate atom-atom interactions with the periodicity of the cavity mode wavelength. We assume that the ratio between the periodicity of the optical lattice and of the cavity mode is incommensurate.}
    \label{Scheme}
\end{figure}

We consider $N$ bosons in a one-dimensional lattice with $K$ sites and open boundary conditions. Their dynamics is described by the extended Bose-Hubbard model with infinite-range interactions, whose Hamiltonian reads:
\begin{equation} \label{eq:hubbard}
    \hat{H} = \hat{H}_\text{BH}+\hat{H}_\text{cav}\,.
\end{equation}
Here, the first term $ \hat{H}_\text{BH}$ describes the standard Bose-Hubbard dynamics with nearest-neighbour hopping at rate $J>0$ and repulsive onsite interaction with amplitude $U$:
\begin{equation}
    \hat{H}_\text{BH} = -J \sum_{i=1}^{K-1}(\hat{b}_i^
    \dagger \hat{b}_{i+1} + \text{c.c.}) + \frac{U}{2}\sum_i^L \hat{n}_i(\hat{n}_i-1)\,,
\end{equation}
where $\hat{b}_i$ and  $\hat{b}_i^\dagger$ are the annihilation and creation operators for a boson on site $i$, and $\hat{n}_i=\hat{b}_i^\dagger\hat{b}_i$ is the number of particle at site $i$. The infinite-range interactions is given by the Hamiltonian \cite{Habibian13}:
\begin{equation}
    \hat{H}_\text{cav} = -\frac{U_1}{K}\qty(\sum_{i=1}^K {\mathcal Z}_\beta(i,\phi_0)\hat{n}_i)^2\,.
\label{Hcav}
\end{equation}
where $U_1$ scales the interaction amplitude, while $\mathcal Z_\beta$ is a function of the site $i$ and of the real parameters $\beta$ and $\phi_0$, that is bounded to unity, $|\mathcal Z_\beta|\le 1$. This function has periodicity $\beta$ in units of the lattice periodicity. In the Appendix we provide its specific dependence on the parameters of the setup of Fig. \ref{Scheme}. $\phi_0$ is an arbitrary phase factor corresponding to different realisations of the effective ``disorder''. In this work we choose $\beta$ to be an irrational number, and in particular $\beta = (\sqrt{5}-1)/2$. As a result Hamiltonian \eqref{eq:hubbard} is quasiperiodic. 

In the rest of this manuscript we assume that $U_1>0$. For this choice, the interaction term favours quasiperiodic density distributions which maximize the expectation value of Eq. \eqref{Hcav} and the quantum ground state exhibits the features of a Bose glass \cite{Habibian13,Habibian13b}. In these works it was shown that Hamiltonian \eqref{eq:hubbard} can be realised in a cavity quantum electrodynamics setup by overlapping the cavity mode to an optical lattice, tightly confining the atoms along the cavity axis. The setup is illustrated in Fig.~\ref{Scheme}. In this configuration parameter $\beta$ is proportional to the ratio between the cavity wave number $k$ and the optical lattice wave number $k_l$, namely, $\beta = k/(2k_l)$.  It is known \cite{Doggen19} that the value of $\beta$ affects the distribution of on-site energies between neighboring sites so the results quantitatively depend on the $\beta$  choice. 

We finally remark that the single-particle ground state and spectrum of these dynamics have been studied in Refs. \cite{Rojan16,Major18}. Here, it was shown that the ground state can be localized for sufficiently large $U_1$. For these values the spectrum exhibits mobility edges.

In the rest of this manuscript we analyse the dynamics for $J=U$ and report the interaction strength $U_1$ in units of $J$.

\section{Spectral properties}

The localisation properties of our system are studied using  exact diagonalisation (ED) of
the Hamiltonian of a system with a unit filling for different number of sites $K$ ranging 
from 7 to 9. These modest {system} sizes are determined by necessity of including the possibility 
to have several bosons at single sites and correspond to dimensions of Hilbert space
between 1716 and 24310.  All quantities 
obtained are also averaged over 500-30000 random $\phi_0$ phases - corresponding to independent
realizations of quasi-periodic disorder. 

\begin{figure}[htbp]
    \centering
    \includegraphics[width=0.9\linewidth]{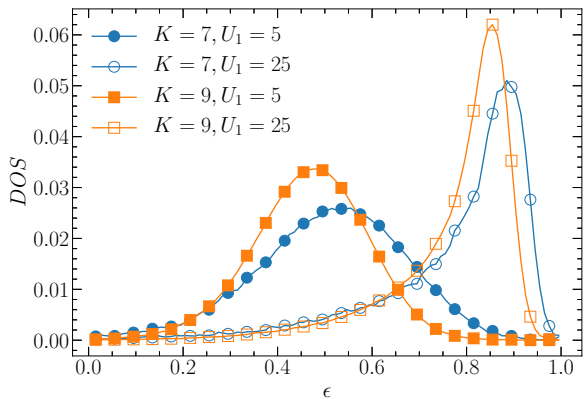}
    \caption{The density of states (DOS) for $K=7, 9$ and $U_1 = 5, 25$ as indicated in the figure.}
    \label{fig:pdf}
\end{figure}

We start our analysis with the density of states (DOS) and its
dependence on the system size, $K$ and cavity mediated interaction strength $U_1$, shown in
Fig.~\ref{fig:pdf}. We rescale eigenenergies $E$  to a unit  interval as
$\epsilon = (E - E_\text{min})/(E_\text{max} - E_\text{min})$ with $E_\text{min}$, $E_\text{max}$
denoting a minimal and maximal energy from each disorder realization, respectively. 
 For moderate value of all-to-all interactions,  $U_1 = 5$,
the maximal DOS {occurs} near the middle of the spectrum. For the higher value $U_1 = 25$ the maximum of DOS is shifted to higher energies, with low and middle energy states laying in the tail of 
DOS. When the system size is increased  the shift of DOS to slightly lower 
energies is observed.
The strong asymmetry of DOS reflects itself in localization properties of eigenstates in different parts
of the spectrum as we show below.

\begin{figure}[htbp]
    \centering
    \includegraphics[width=0.49\linewidth]{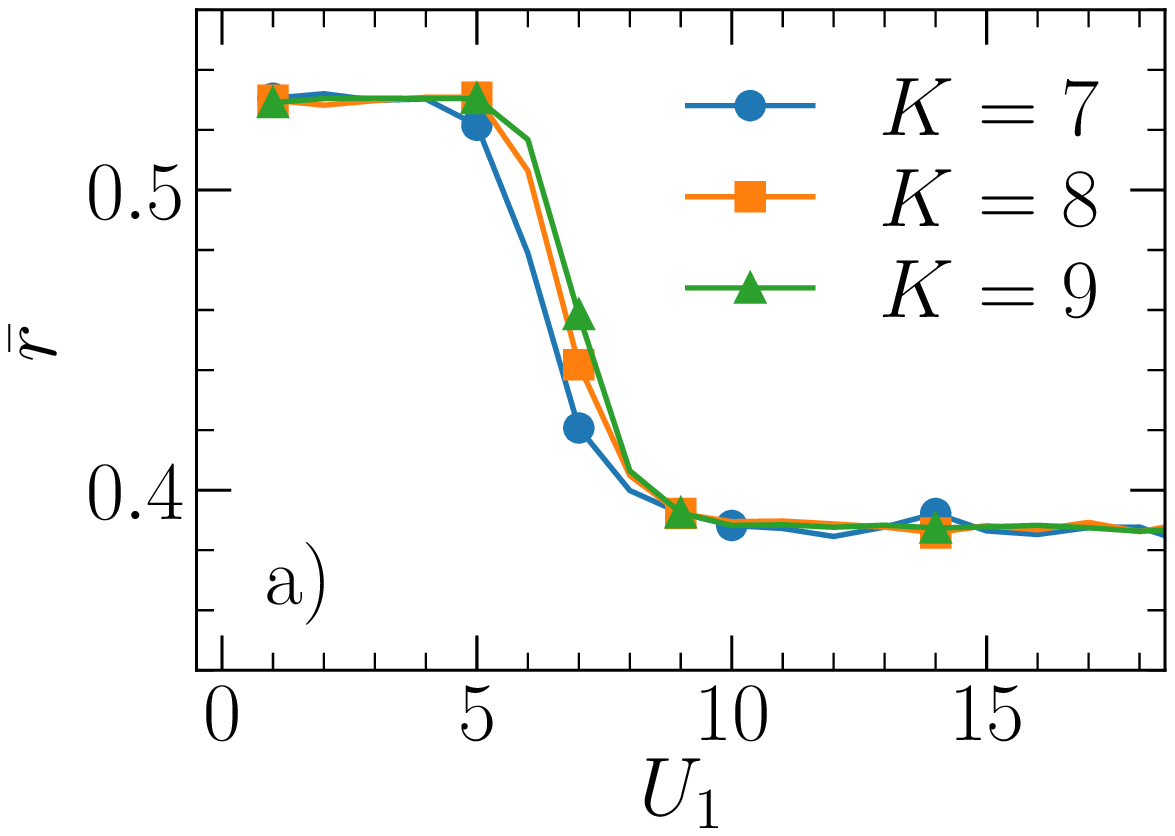}
    \includegraphics[width=0.49\linewidth]{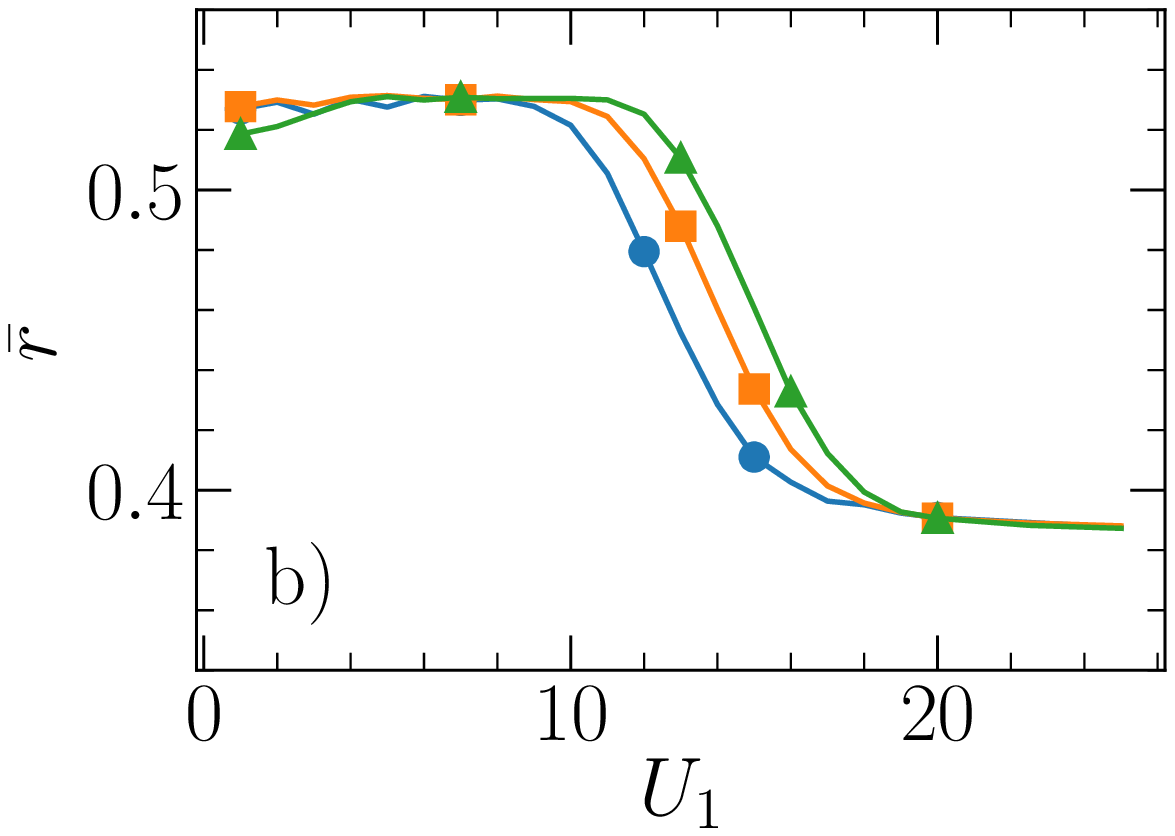}
    \caption{The dependence of the mean gap ratio $\bar{r}$ on the all-to-all interaction strength
    $U_1$ for different energy ranges: a) $\epsilon \in [0.3, 0.4]$, b) $\epsilon \in [0.6, 0.7]$.}
    \label{fig:rbar_U1}
\end{figure}

To study the localization properties of eigenstates  we utilize the 
mean gap ratio \cite{Oganesyan07}. It is defined as
\begin{equation}
    \bar{r} = \left<\min\qty{\frac{\epsilon_{i+1}-\epsilon_i}{\epsilon_i-\epsilon_{i-1}},\frac{\epsilon_i-\epsilon_{i-1}}{\epsilon_{i+1}-\epsilon_i}}\right>,
\end{equation}
where the averaging is performed in a band of eigenenergies around a given rescaled energy $\epsilon$ and
over disorder realizations. For {ergodic system, with level statistics
well described by Gaussian Orthogonal Ensemble of random matrices,} $\bar{r} \approx 0.53$
while $\bar{r} \approx 0.39$ {indicates} localisation and Poissonian {level} statistics \cite{Atas13}. 
The dependence $\bar{r}$ on the interaction strength  {$U_1$}
for different parts of spectrum and system sizes is shown in Fig.\ref{fig:rbar_U1}. 
The $\bar{r}$ plots show a couple of interesting properties of our system. Firstly,
for a fixed system size and energy range, the {system} undergoes a crossover between ergodic and localized regimes
as we increase $U_1$, meaning that stronger interactions actually prohibit thermalization. 
Thus, the interaction induced randomness to some extent plays a role of a
(quasi)random on-site potential in the interacting system. A similar
phenomenon occurs for
 bosons interacting by random contact interactions \cite{Sierant17,Sierant17b,Sierant18}. Here,
we keep the uniform on-site interactions fixed at low $U=1$ value and the randomness comes from infinite-ranged
cavity mediated interactions only. 
A second observation is that the higher the energy is, the stronger $U_1$ is necessary to induce
the crossover to localized phase. Therefore, for a fixed $U_1$, we observe a so-called mobility
edge, namely the lower energetically part of the spectrum is localized, while the higher is thermal
(delocalized). However, Fig.~\ref{fig:rbar_U1} b) shows that the value of $U_1$
at which the system enters into the localized regime increases with system size $K$, implying that the mobility edge
shifts to lower energies with  increasing $K$. This suggests
that the mobility edge may not survive in the thermodynamic limit and in 
this limit all eigenstates are extended.

\begin{figure}[htbp]
    \centering
    \includegraphics[width=0.49\linewidth]{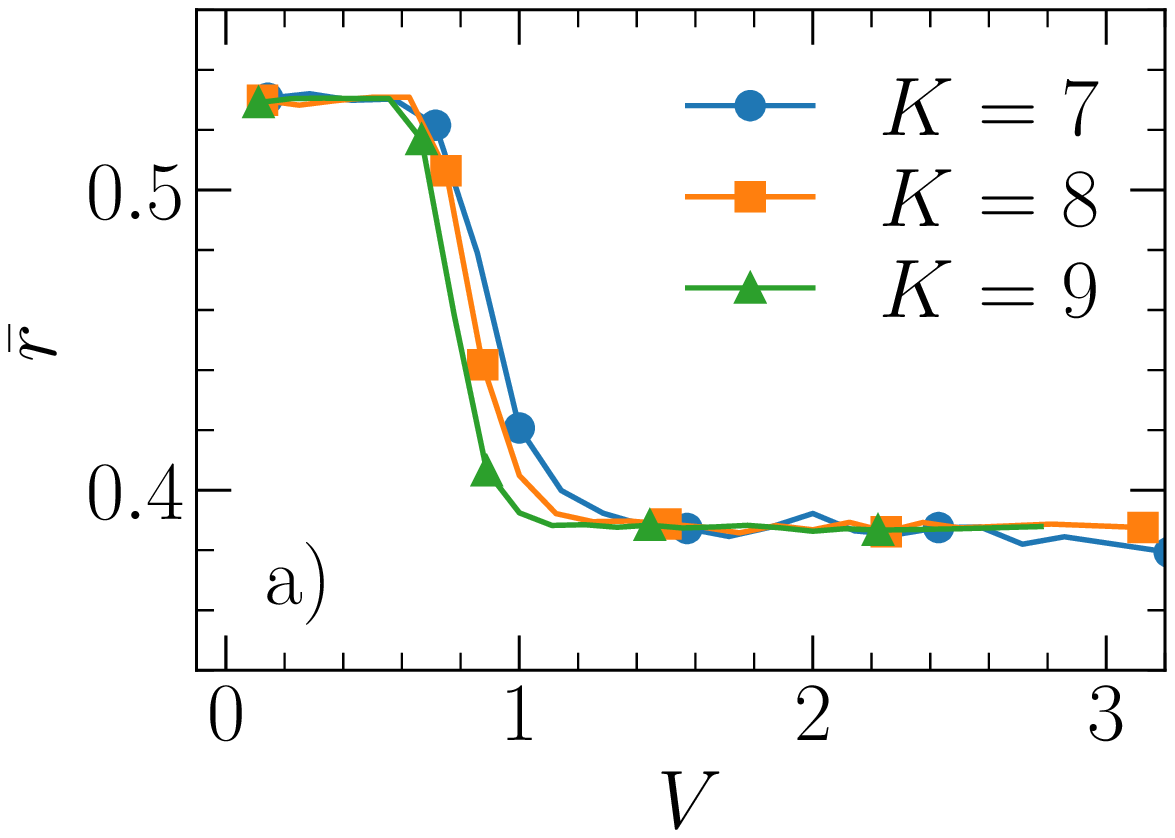}
    \includegraphics[width=0.49\linewidth]{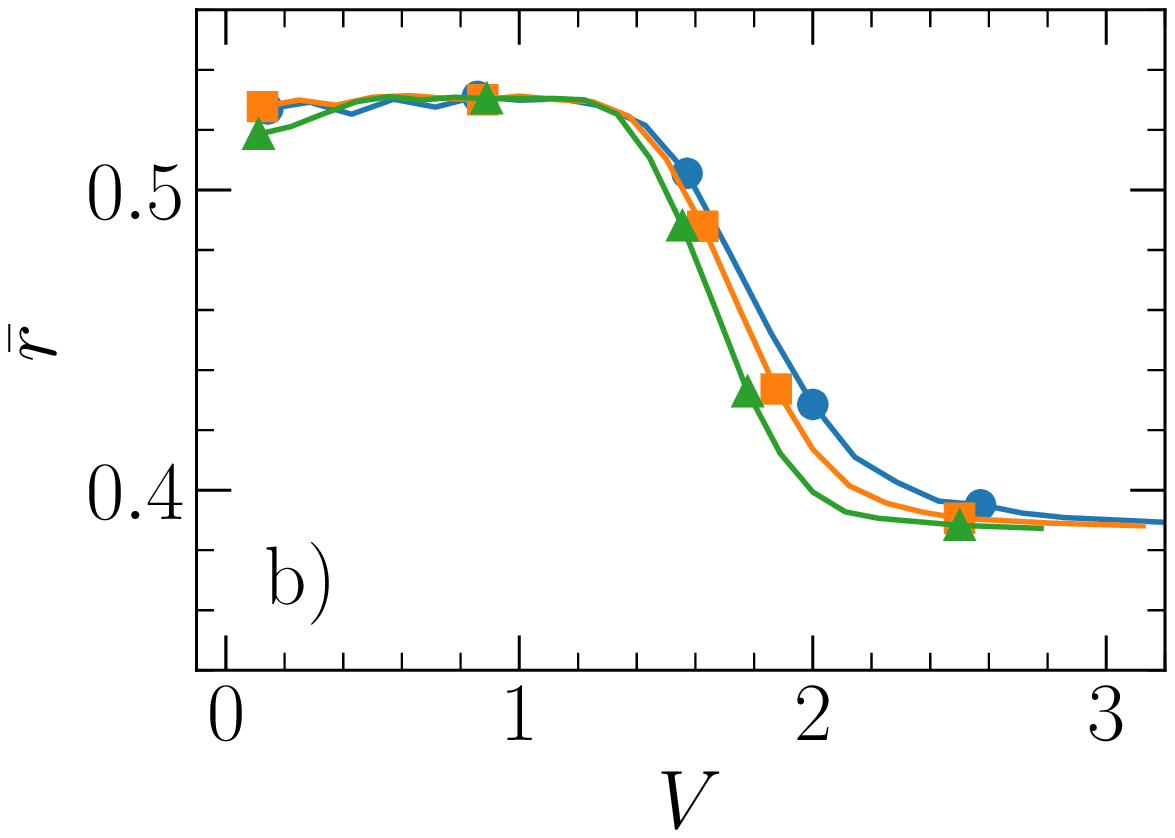}
    \caption{The dependence of $\bar{r}$ on $V = U_1/K$ for various energy ranges. a) $\epsilon \in [0.3, 0.4]$, b) $\epsilon \in [0.6, 0.7]$.}
    \label{fig:rbar_V}
\end{figure}

Observe,
however, that we can choose a different scaling the coupling strength in
$\hat{H}_\text{cav}^{(0)}$ term, \eqref{Hcav}. The factor in front of the sum{,} $U_1/K${,}
has been chosen, in accordance with the customary approach
\cite{Habibian13,Dogra16}  that ensures the appropriate extensive scaling of the Hamiltonian with increasing $K$.
In fact, it may be justified by assuming that the  increase of the system size $K$,
for a fixed wavenumber $k$, implies the increase in 
the cavity length and thus the appropriate scaling of the cavity mode {volume}. However, another approach is possible -
to keep the cavity length fixed and change the lattice spacing. 
Then, instead of the coefficient $U_1/K$ in front of the all-to-all interactions term, we
would have a constant $V$} that remains fixed {when} $K$ {is increased}. Such a scaling implies a nonstandard thermodynamic limit
as the energy becomes a super extensive quantity. A similar
procedure was adopted in \cite{Gopalakrishnan19} {to study MBL in presence of power-law interactions}. 
{We show the mean gap ratio}  $\bar{r}$ 
as a function of $V$ in Fig.~\ref{fig:rbar_V}. 
For a sufficiently large $V$ a crossover to
{localized regime} is observed. 
However, contrary to fixed $U_1$ case, the resulting $\overline r(W)$ curves 
deviate from the ergodic value  $\overline r\approx0.53$ at a system size independent value of $V$ and 
the crossover becomes
steeper with increasing system size. 
This suggests that the observed crossover
{becomes a sharp transition to an MBL phase at a fixed critical value of $V$} 
{in the considered non-standard} thermodynamic limit. 
We note that a similar behavior at the ergodic-MBL crossover was observed for disordered $U(1)$ lattice
gauge theory \cite{Giudici20}, in which the elimination of gauge field leads to long-range interactions.

Let us note that identifying the transition for $V=const.$ limit implies the delocalized limit for $U_1=const.$ 
for arbitrary $U_1$. On the other hand all experiments are performed for a finite $K$ value that may reach several
tens or hundreds but necessarily remains finite so our observations for the standard $U_1=const$
scenario are {experimentally relevant}. 

\begin{figure}[htbp]
    \centering
    \includegraphics[width=0.49\linewidth]{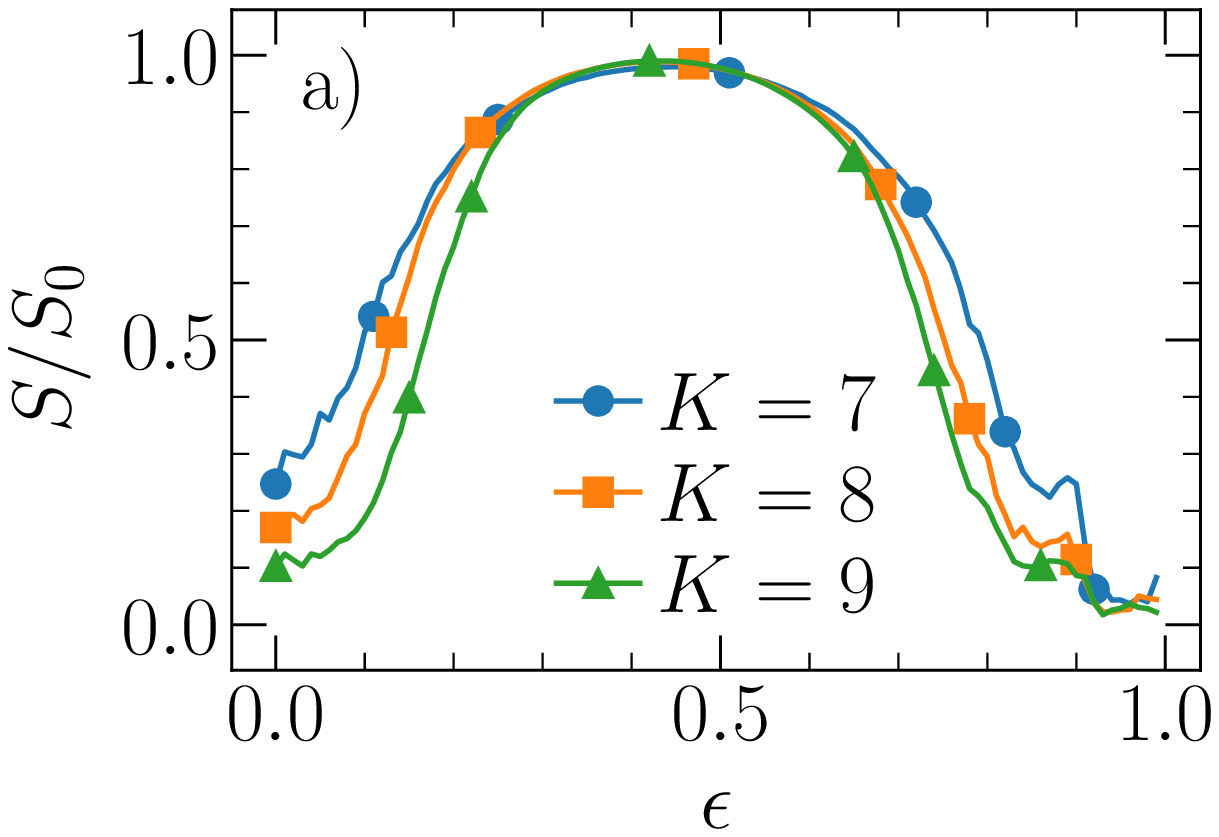}
    \includegraphics[width=0.49\linewidth]{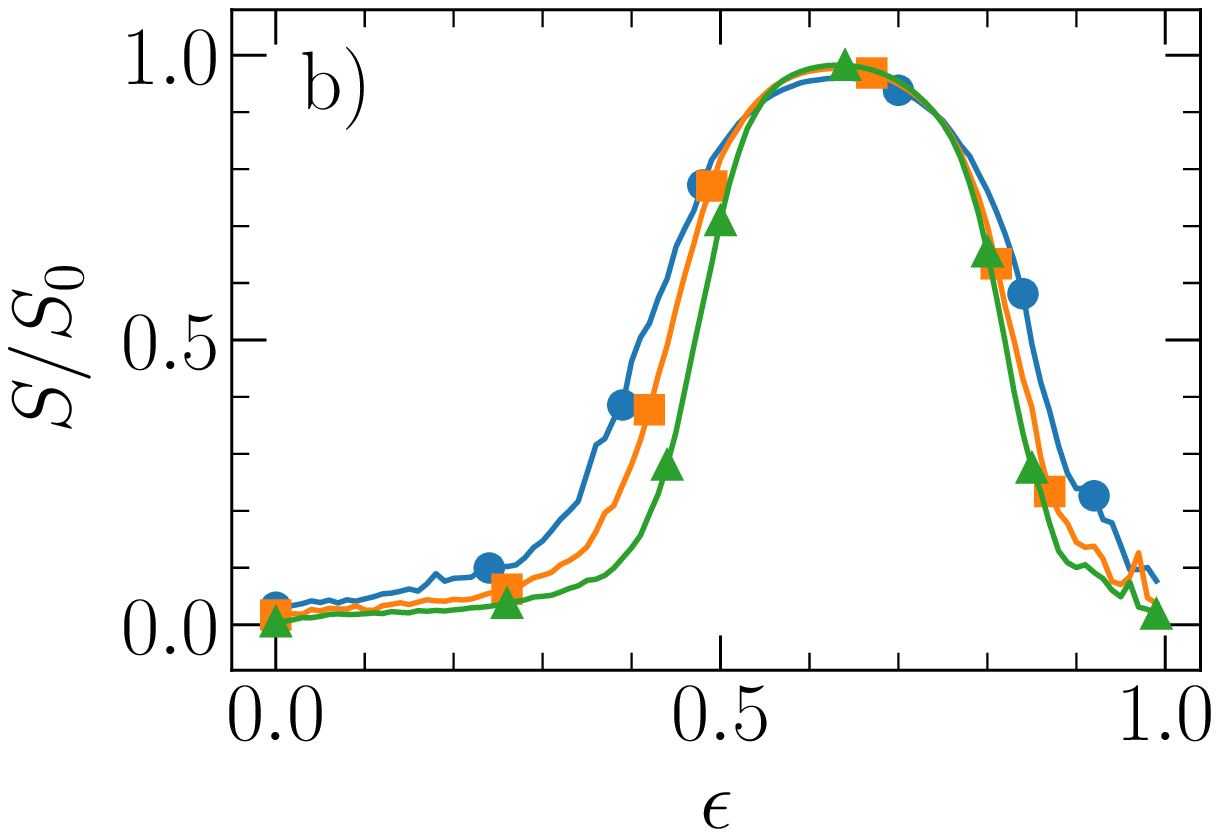}
    \includegraphics[width=0.49\linewidth]{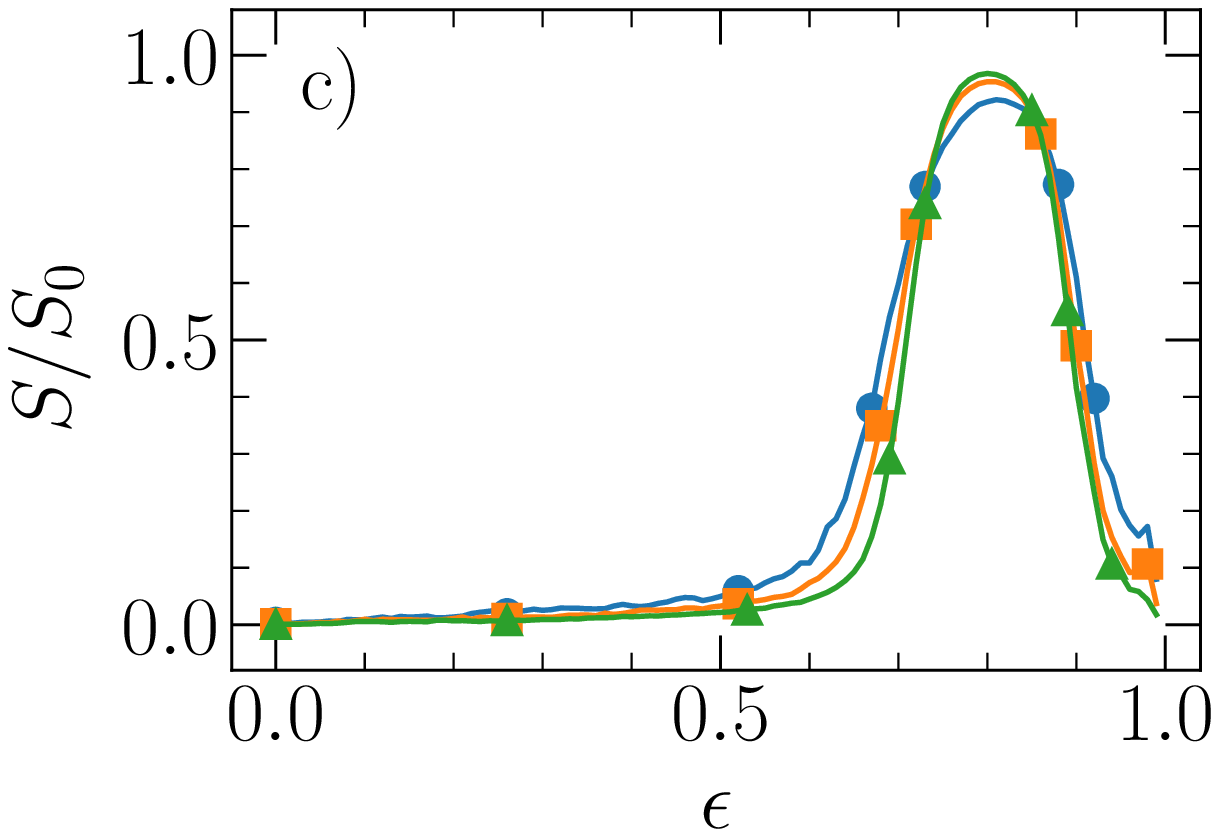}
    \includegraphics[width=0.49\linewidth]{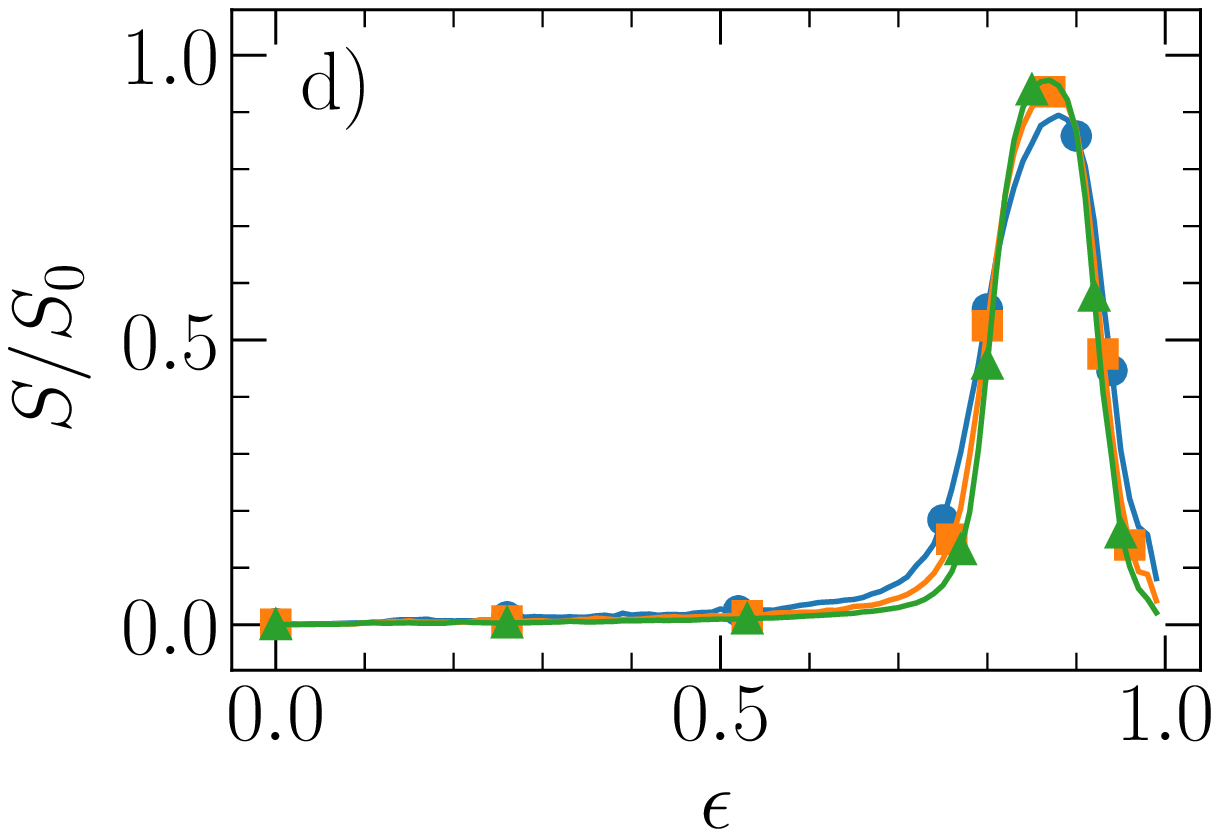}
    \caption{The $\epsilon$ dependence of von Neumann biparite entanglement entropy $S{/S_0}$ 
    of eigenstates normalized by a random Gaussian state value {$S_0$}. 
    Subsequent panels correspond to different values of $V = U_1/K$: a) $V = 0.5$, 
    b) $V = 1$, c) $V = 2$, d) $V = 3$.}
    \label{fig:eig_S_V}
\end{figure}

The crucial difference between extended and localized states can be seen in their entanglement properties.
For extended states quantum entanglement between subsystems which we call
$A$ and $B$ resulting from splitting 
of the entire system into two, {follows a}
volume law, while 
for localized states area law is expected. To quantify the entanglement, we calculate the von Neumann biparite entanglement entropy defined as
\begin{equation} \label{eq:biparite}
    S = -\Tr_B[\hat{\rho}_B \ln \hat{\rho}_B],
\end{equation}
where $\hat{\rho}_B=\Tr_A[\hat{\rho}_{AB}]$ is a (usually  mixed) state of the subsystem $B$ obtained by 
tracing the state $\hat{\rho}_{AB}=\ket{\psi}\bra{\psi}$ 
over degrees of freedom of complementary subsystem $A$.
If subsystems $A$ and $B$ are not entangled, $\hat{\rho}_B$ is a pure state and $S = 0$.
Fig.~\ref{fig:eig_S_V} presents the entanglement entropy of eigenstates ($\hat{\rho}_{AB}=\ket{\psi_n}\bra{\psi_n}$,
where $\hat{H}\ket{\psi}=E_n \psi_n$) 
{averaged over eigenstates with energies corresponding to the rescaled energy $\epsilon$}
for different values of $V$ and $K$, normalized by the entropy $S_0$ of random Gaussian states {\cite{Vidmar17}}.
In this case, $A$ and $B$ subsystems are two equal $K/2$ {parts}.

Again we observe two main features. Firstly, for low lying energy states the normalized entropy is close to 0
and independent of system size $K$, 
indicating  localized states. However
for higher energies $\epsilon$, the entanglement entropy
$S$ approaches $S_0$  as expected for thermal states. 
Interestingly, for very high energies we observe the decrease in $S$, but this appears at the edge of 
the spectrum where the density of states is again very low. Secondly, increasing $V$ shifts the the thermal region to higher energies, which is consistent with level statistics. 
What is important, for constant 
$V$ and changing $K$ we again observe the crossing of curves corresponding to different system sizes. 
Interestingly, the crossing is visible on both sides of $S$ maximum, {suggesting that the localized eigenstates, fulfilling the area law, found at the} highest
energies $\epsilon$ can also survive in the non-standard thermodynamic limit.

\begin{figure}[htbp]
    \centering
    \includegraphics[width=0.9\linewidth]{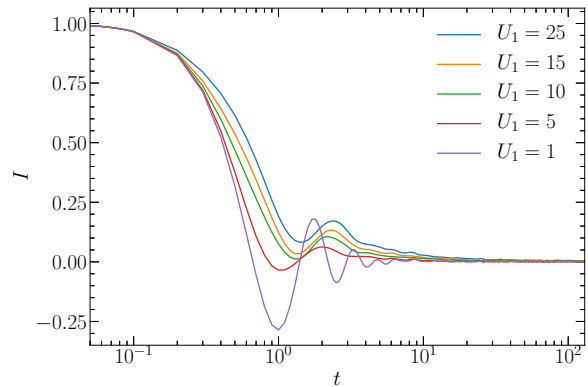}
    \caption{Population imbalance $I(t)$ during time evolution of a density-wave like state for $K = 8$ and a several of $U_1$ values: 25, 15, 10, 5, 1 respectively, starting from the top.}
    \label{fig:imbalance}
\end{figure}

\section{Time dynamics}
In the previous Section we analysed spectral properties discussing various aspects of
the crossover between ergodic and MBL regimes
as well as the presence of the mobility edge. Here, we address 
more experimentally relevant issue - whether the localization properties may be observed 
in the time dynamics starting from a judiciously prepared initial state.
Following the usual experimental strategy \cite{Choi16,Luschen17}
we might consider the population imbalance, defined as
\begin{equation}
    I(t) = \left<\frac{1}{K}\sum_{i=1}^{K}(-1)^i\hat{n}_i(t)\right>.
\end{equation}
{An initial state with an imprinted density pattern, for which the imbalance $I(t)$ 
is non-zero at $t=0$ can be used to demonstrate ergodicity breaking.}
For unit density assumed by us {such a state can be chosen as} 
$\ket{2020\dots}$. The system evolves governed by $\hat{H}$ and the $I(t)$ is monitored to reveal 
the system properties. The initial unit imbalance decays to 0 for systems obeying the 
eigenstate thermalization hypothesis \cite{Deutsch91,Srednicki94}. In such a case the  information about
the initial configuration is lost during the course of time evolution. 
In contrast, a non-zero long-time value of $I$ suggests that the information is
partially preserved and is indicative of the localized phase. The evolution of the population
imbalance for a several values of $U_1$ is shown in Fig.~\ref{fig:imbalance}. The results are 
obtained using Chebyshev propagation \cite{Tal-Ezer84,Fehske08} relying on sparse matrix 
operations, which allows us to reach system sizes up to $K=12$ with reasonable computational
time and resources. Surprisingly, although the level statistics (Figs.~\ref{fig:rbar_U1},\ref{fig:rbar_V}) 
indicate localization in some ranges of $U_1$, for all parameters studied here, the imbalance
for the initial density wave state rapidly decays to zero suggesting thermalization. 
It is easily understood after inspecting the energy of the density wave state {$\ket{2020...}$} (energy is
conserved during propagation) which, for all reasonable choices
of the parameters of the {problem}, corresponds
to regions {identified as ergodic by the mean gap ratio}. 

\begin{figure}[htbp]
    \centering
    \includegraphics[width=0.49\linewidth]{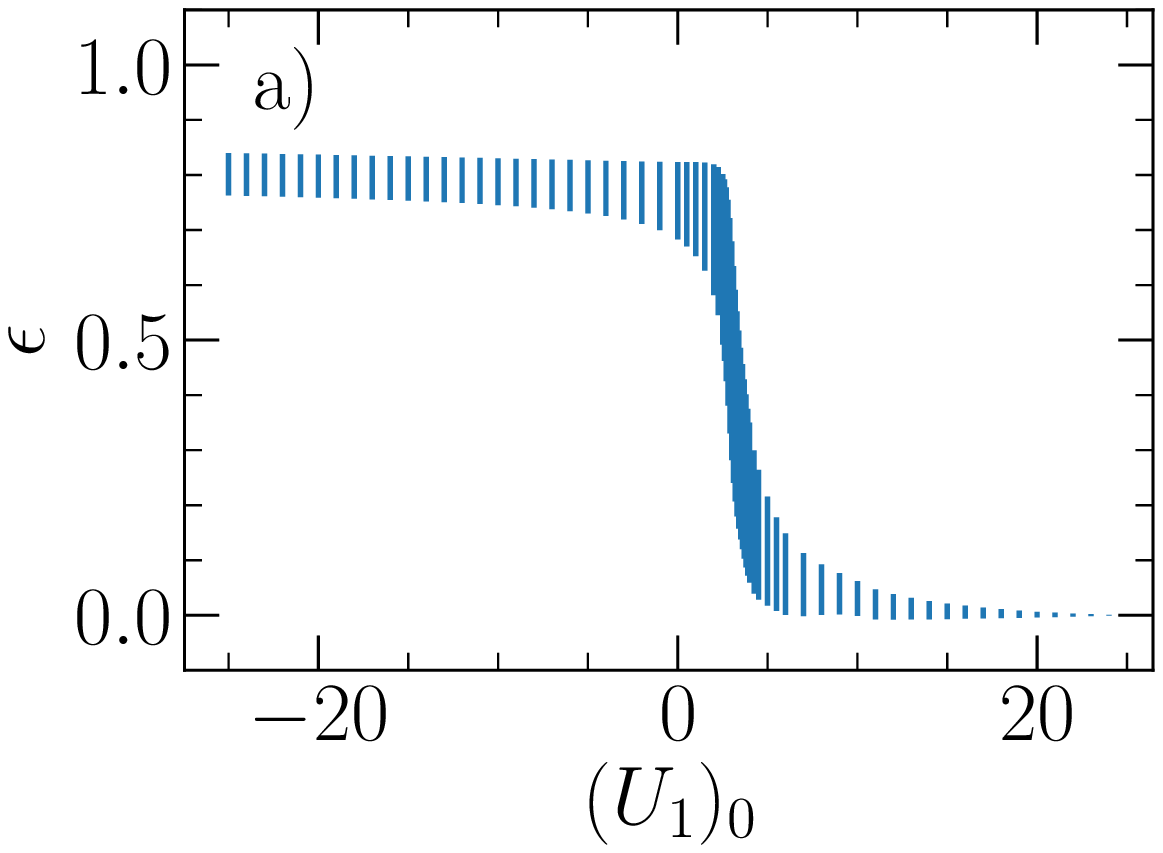}
    \includegraphics[width=0.49\linewidth]{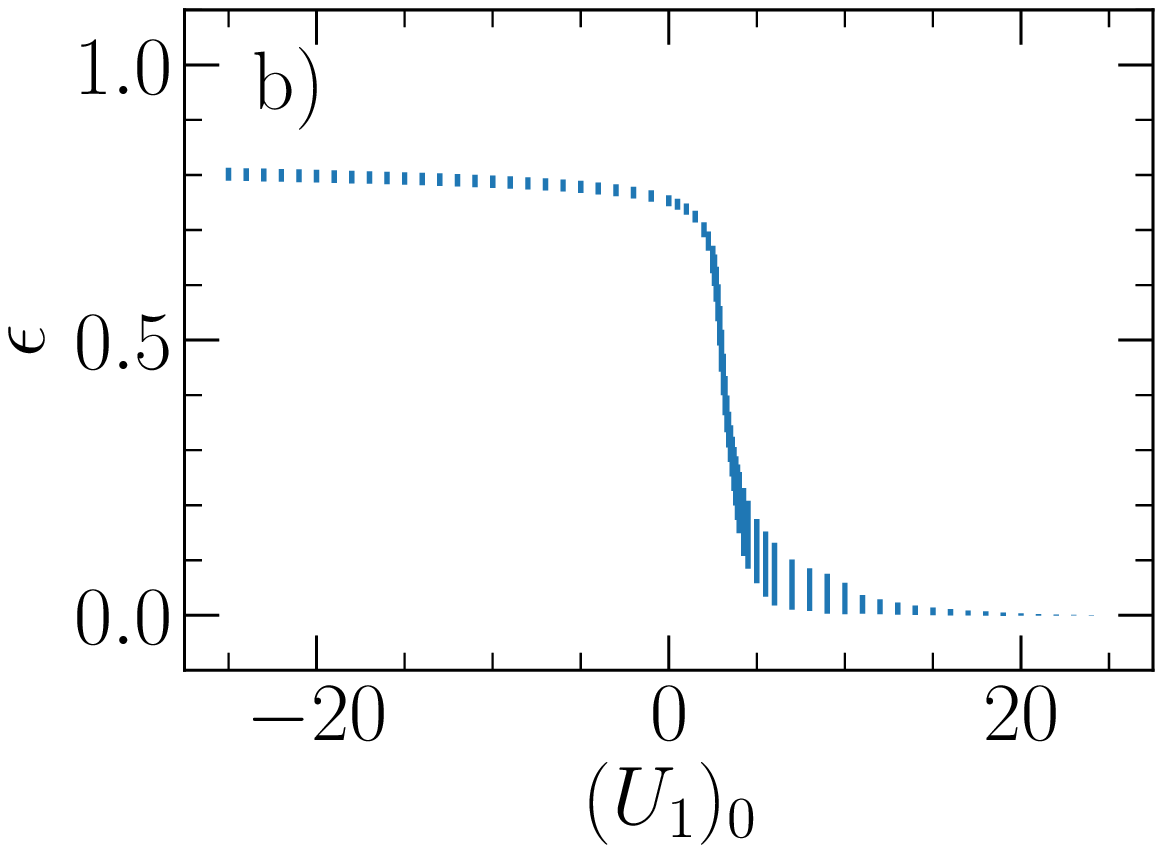}
    \includegraphics[width=0.49\linewidth]{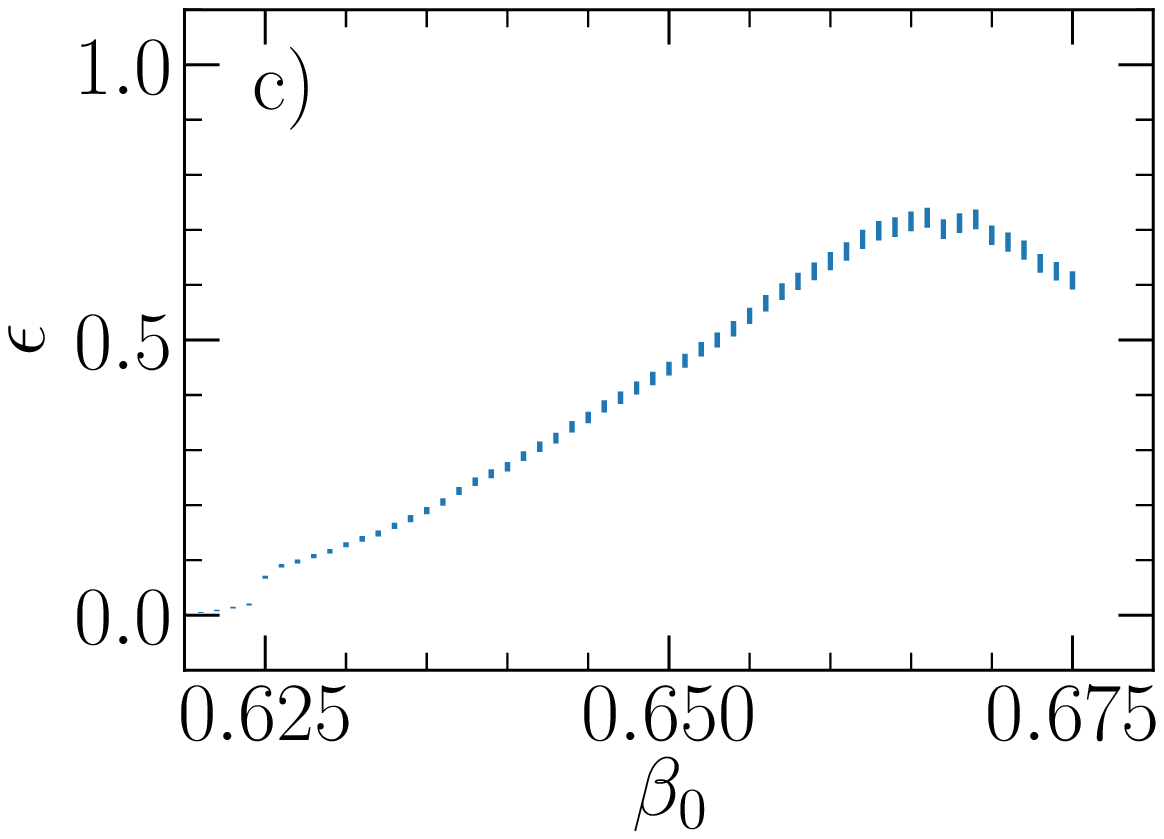}
    \includegraphics[width=0.49\linewidth]{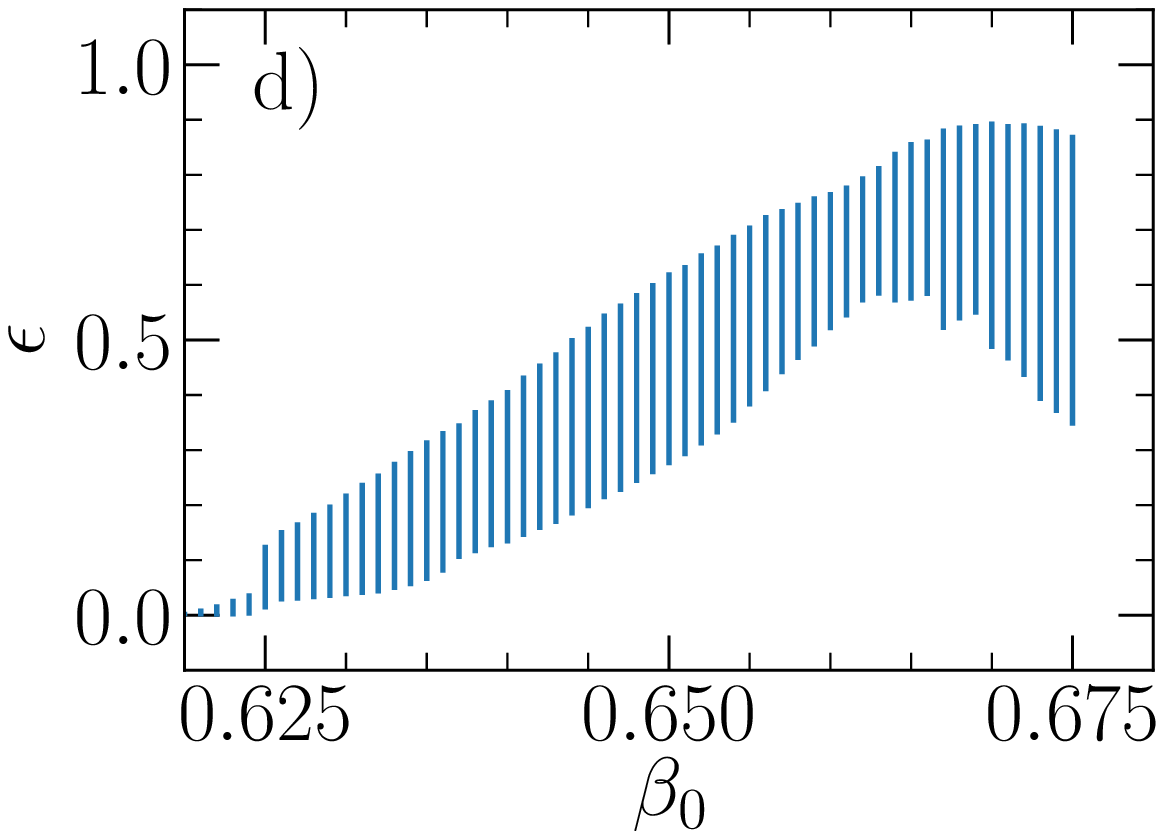}
    \includegraphics[width=0.49\linewidth]{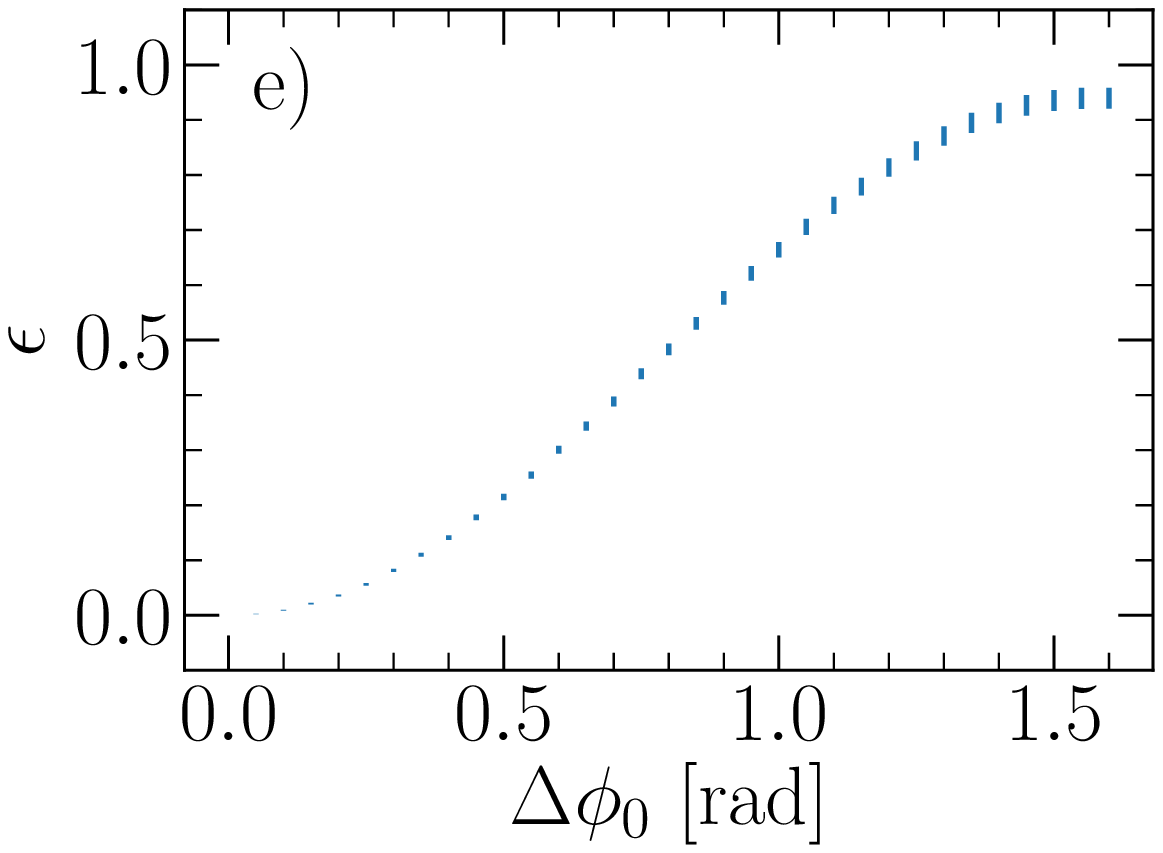}
    \includegraphics[width=0.49\linewidth]{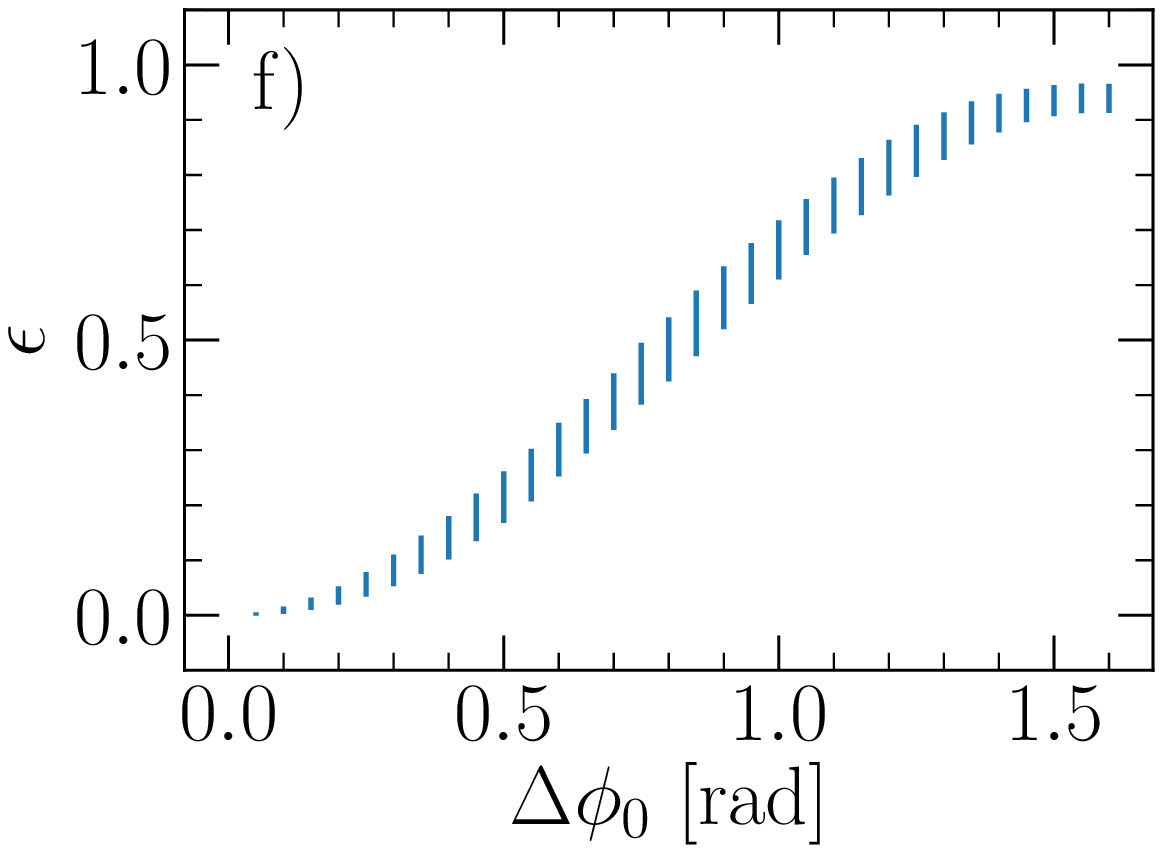}
    \caption{The dependence of average energy $\epsilon$ of the quenched state w.r.t. the final Hamiltonian on several quenched parameters for $K = 8$. On both panels in each row the same energy values are presented, but with different types of errors represented by vertical lines. In the left column they represent the quantum error $\sqrt{\expval{\hat{H}_\text{final}^2}-\expval{\hat{H}_\text{final}}^2}$, while in the right column -- the standard deviation of the random phase distribution of mean energies.}
    \label{fig:quench}
\end{figure}

One can imagine a preparation of a different initial product state, e.g. a uniformly
occupied state $ \ket{1111\dots}$ and use the transport distance \cite{Rispoli19} 
(see also below) to look for signatures of localization. Our attempts in that direction failed (except for 
highly unusual states
which did not allow us to study the breaking of ergodicity
in our system in a systematic manner).
The reason for that is simple. To reach the MBL regime in our system we need a sufficiently strong disorder which 
appears solely in the cavity mediated interaction term. For large $U_1$ this term dominates the energy of
product states. Thus, the occupation numbers of an initial product state
must be adapted in such a way that the all-to-all interaction term is possibly small {in order to reach localized regime}.

To inspect the low energy regime in a more systematic manner 
we employ an 
 alternative approach  introduced by \cite{Naldesi16} and termed the quantum quench spectroscopy (QQS). 
 In the first step {of QQS} the system is prepared in the ground state of some initial Hamiltonian $\hat{H}_\text{ini}$.
 An abrupt change of one of its parameters to the desired  $\hat{H}$ follows. In such a procedure the initial 
 state is not changed but now becomes a non{-}stationary initial superposition of eigenstates of $\hat{H}$. 
 One may hope 
 that an appropriate choice {of the 
 initial Hamiltonian $\hat{H}_\text{ini}$} may allow to probe different
 parts of the spectrum of the final Hamiltonian {$\hat{H}$}. It was shown in \cite{Naldesi16}
 that this procedure, {when} ramping the disorder strength, may allow {one} to probe the mobility edge in the system of interacting 
 spinless fermions in a quasiperiodic potential.
 
 It is important, however to realize two possible limitations. As the initial state is not
 an eigenstate of $\hat{H}$ its energy suffers from quantum uncertainty
 \begin{equation}
     \Delta E= \sqrt{\expval{\hat{H}^2}-\expval{\hat{H}}^2},
 \end{equation}
 the fate shared, in fact, also by the initial product states. 
  Moreover, as a target mean energy depends on a random $\phi_0$ phase in \eqref{Hcav}, it is also a random variable, 
  whose dispersion can be characterized by a sample standard deviation. This second  uncertainty is
  known to decrease with an increasing system size \cite{Naldesi16}, so it is not a major concern 
  in the thermodynamic limit. On the other hand it may be an obstacle for relatively small system sizes
  treated by ED in this work.  

Comparison of mean energies obtained and both uncertainties is shown in Fig.~\ref{fig:quench} 
for different possible quenche scenarios. 
We fix $\beta=(\sqrt{5}-1)/2 \approx 0.618$ and $U_1 = 25$ as the final values. 
In the first approach we quench the disorder amplitude from some initial
 $(U_1)_0$ value to $U_1$ - the path similar to \cite{Naldesi16}. The results presented in the upper row of Fig.~\ref{fig:quench} 
 show low energy selectivity. For small difference between the initial and final disorder amplitude low excitations are reached only.
 The rapid jump in relative energy $\epsilon$ appears for low initial interaction strength.
 When $(U_1)_0$ has a different sign than $U_1$ a relatively large $\epsilon$ values may be reached but with quite
 significant quantum uncertainty. Similarly, the region of small $(U_1)_00$ where a rapid change of final
 energies is observed suffers from that uncertainty. 
 
 Situation is entirely different if, instead of the interaction strength, the ratio of cavity and laser
 wavenumbers is changed, i.e. $\beta$ as shown in the middle row of Fig.~\ref{fig:quench}. The quantum uncertainties
 are small and a broad range of final energies may be achieved, however $\phi_0$ averaging uncertainty is 
 very large for the system sizes under consideration. Moreover, the maximal energy achieved is
 around $\epsilon \approx 0.70$, which is below the mobility edge estimated for $K=8$ from gap ratio 
 as $\epsilon \approx 0.75-0.80$ not enabling us to probe the  crossover between ergodic and 
 MBL regimes. 
 
 Finally, the bottom row in Fig.~\ref{fig:quench} shows yet another possible quench. For each (random)
 $\phi_0$ we choose a different $\phi_0(0)$ and define the quench as a rapid change of the phase from 
 $\phi_0(0)$ to $\phi_0$. We denote the magnitude of the phase change by $\Delta\phi_0$. As observed 
 the quantum and statistical uncertainties are relatively small. Moreover,  the final energy approaches 
 1 for $\Delta\phi_0 = \pi/2$. Thus, we perform further the {QQS} simulations using {this} type of quench.

\begin{figure}[htbp]
    \centering
    \includegraphics[width=0.5\linewidth]{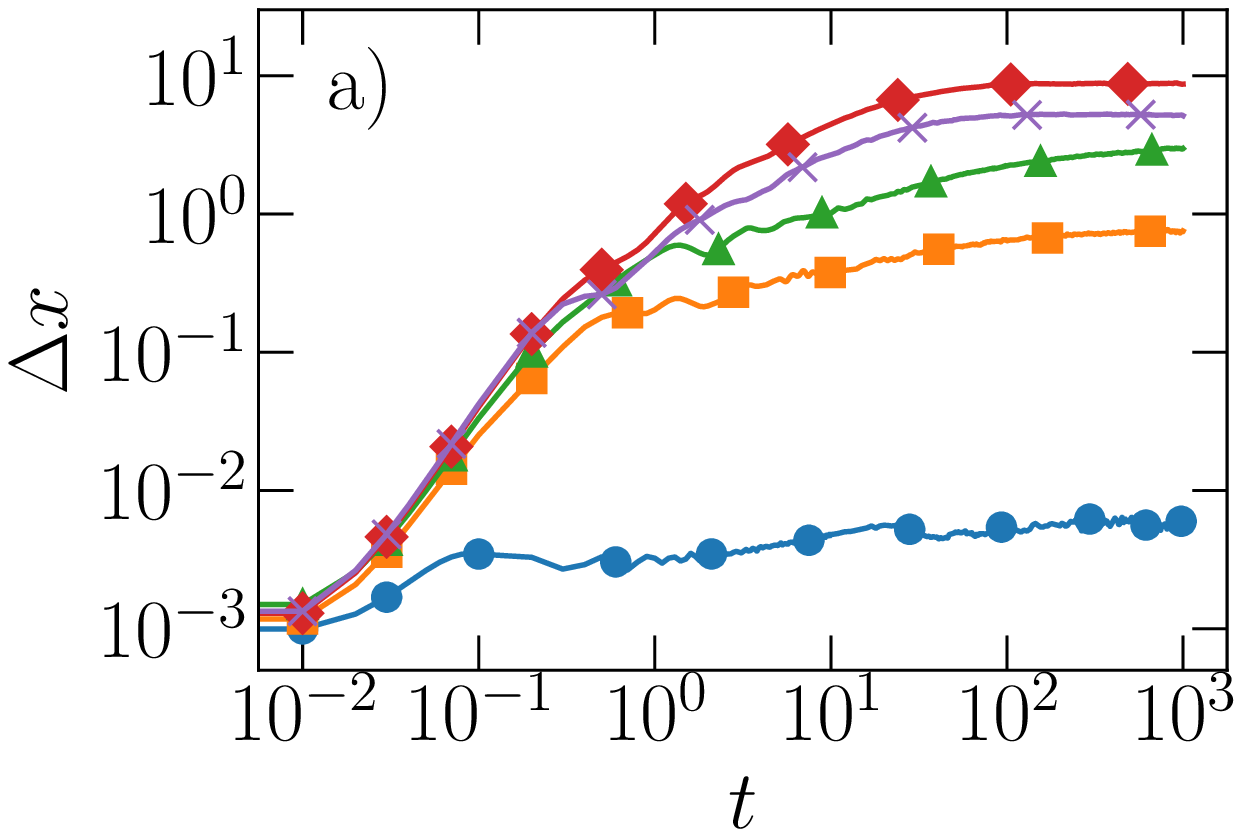}
    \includegraphics[width=0.48\linewidth]{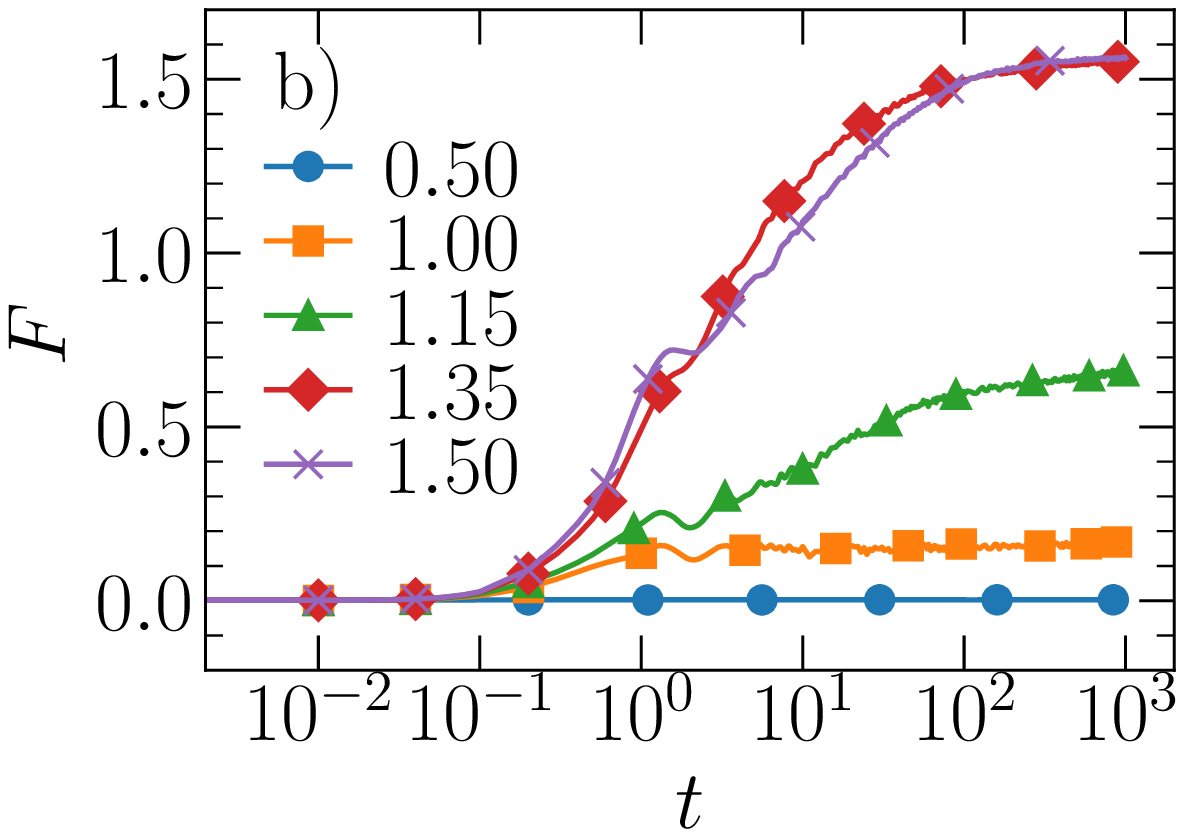}
    \includegraphics[width=0.485\linewidth]{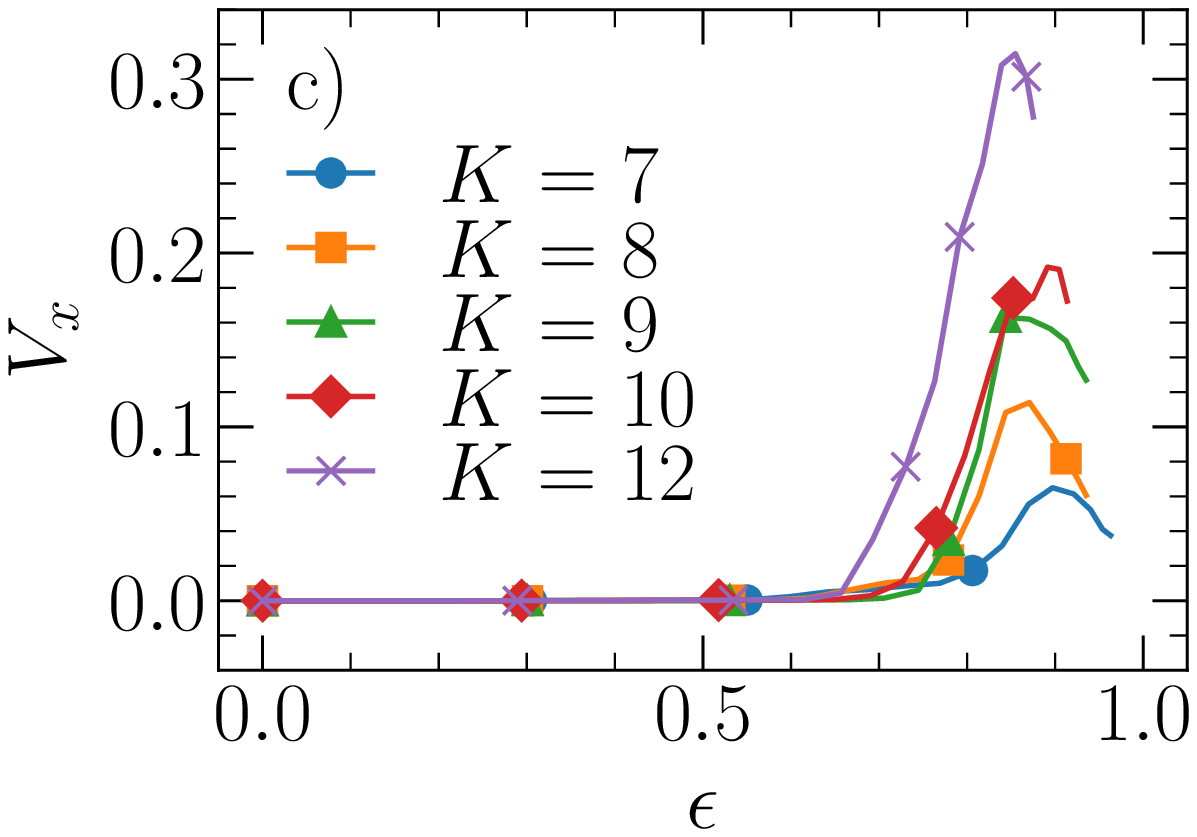}
    \includegraphics[width=0.495\linewidth]{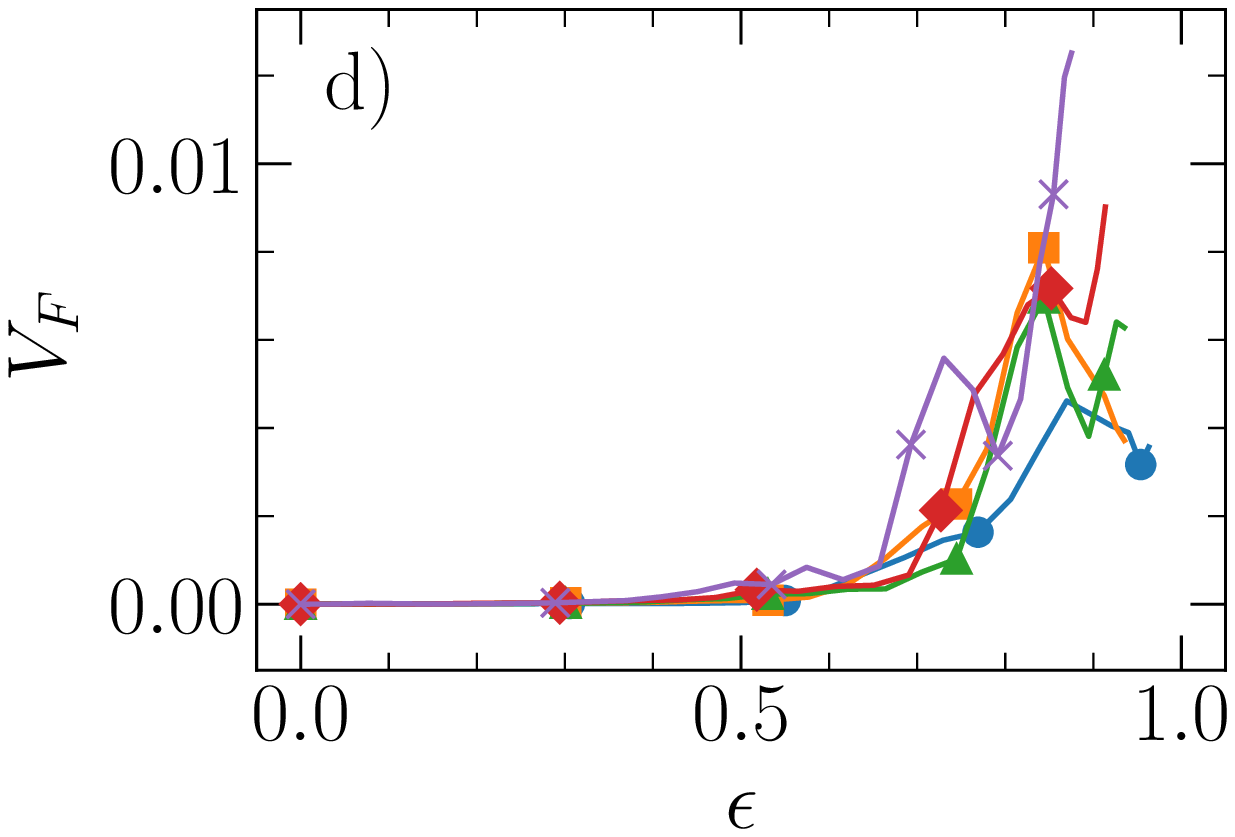}
    \includegraphics[width=0.50\linewidth]{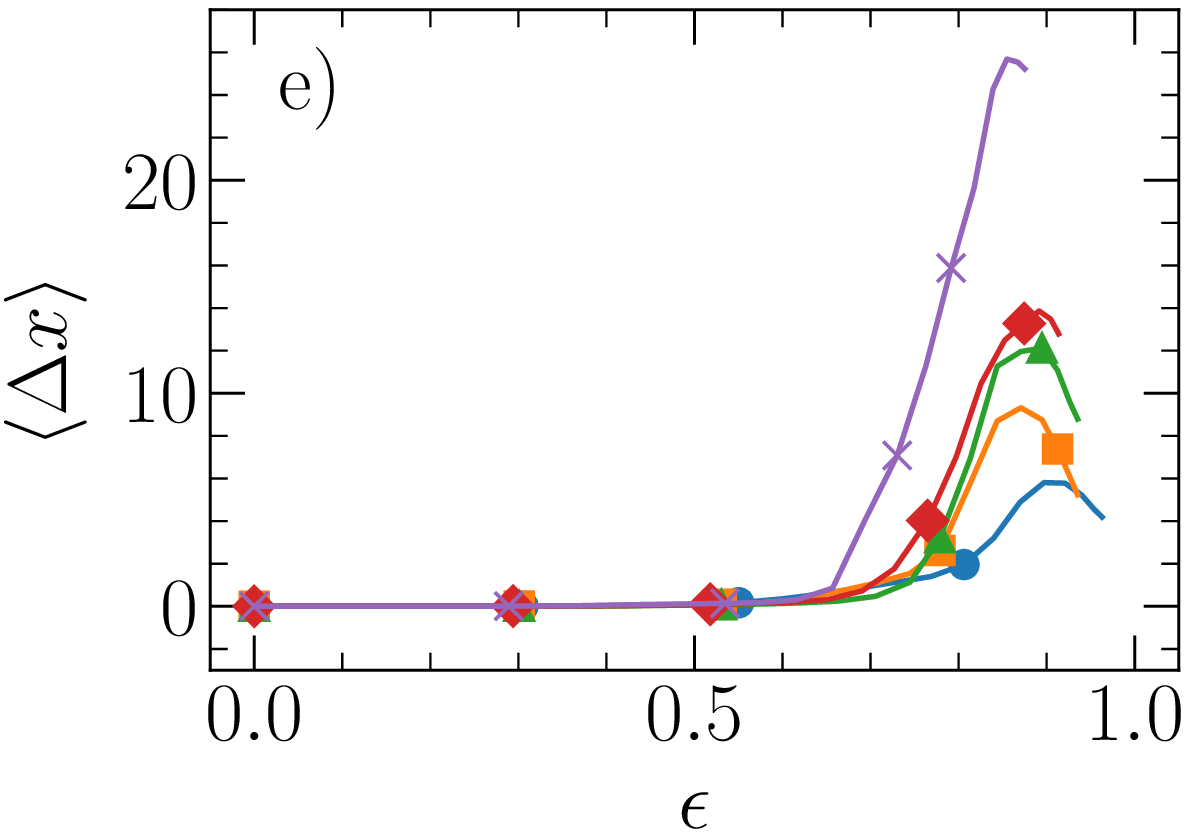}
    \includegraphics[width=0.48\linewidth]{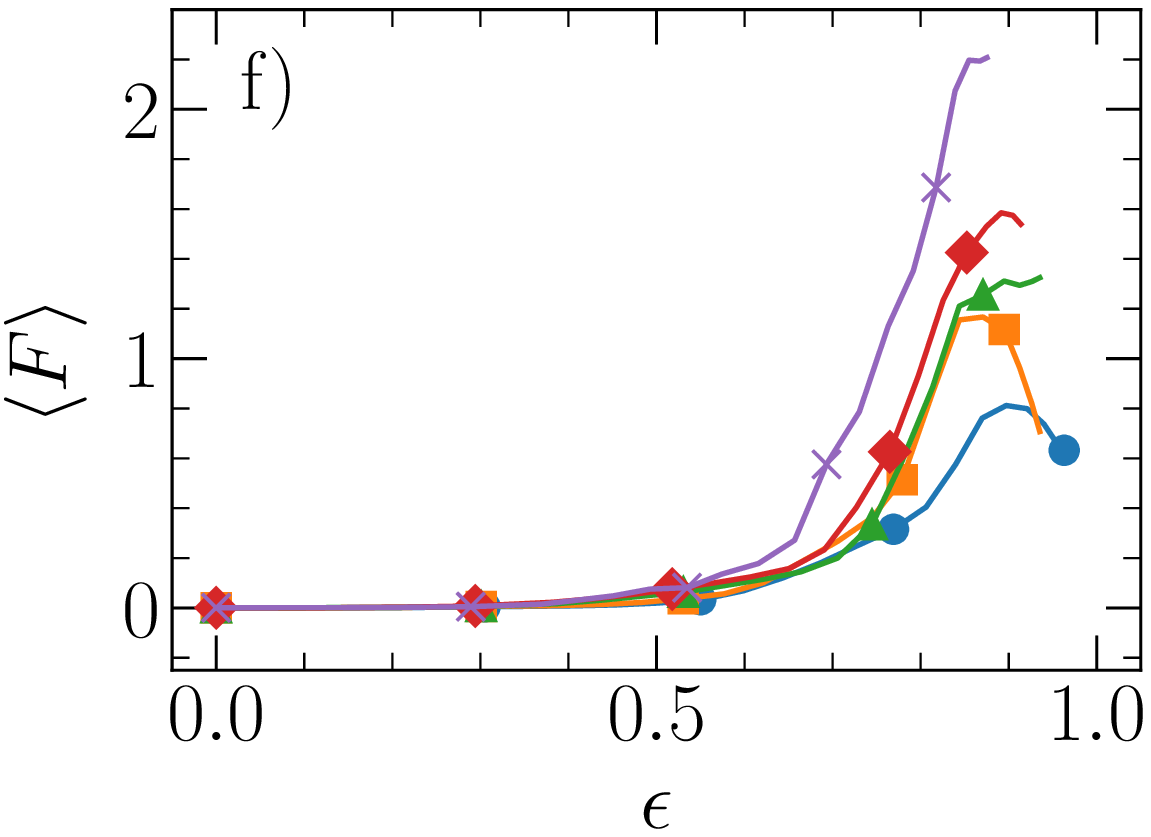}
    \caption{a) Time evolution of transport distance for $K = 8$ and a several $\Delta\phi_0$ values. b) A similar evolution of fluctuations $F$. c) Maximal transport pseudo-velocity dependence on initial state energy for a several $K$. d) Fluctuations pseudo-velocity energy dependence. e) Average value of $\Delta x$ for $t \in [100, 500]$ energy dependence. f) Average value of $\Delta x$ for the same time interval energy dependence.}
    \label{fig:dist_fluct_evolution}
\end{figure}

After a change of the cavity wavenumber by $\Delta\phi_0$
we consider time evolution of the system. Instead of the Imbalance  we calculate, following \cite{Rispoli19} 
the transport distance,  defined as
\begin{equation}
    \Delta x = 2\abs{\sum_{d=1}^{K-1} d \times \overline{\expval{G_c^{(2)}(i,i+d)}_i}},
\end{equation}
where $G_c^{(2)}(i,i+d)=\expval{\hat{n}_i\hat{n}_{i+d}}-\expval{\hat{n}_i}\expval{\hat{n}_{i+d}}$ and 
averaging over sites $\expval{\dots}_i$ is done before averaging over
realizations $\overline{(\dots)}$. Note that the measurement of $\Delta x$ requires 
access to two-site correlations available in state of the art experiments \cite{Rispoli19}. 
The   second quantity we calculate is {an} edge fluctuation defined as
\begin{equation}
    F = \frac{\qty(\expval{\hat{n}_1^2} - \expval{\hat{n}_1}^2) + \qty(\expval{\hat{n}_K^2} - \expval{\hat{n}_K}^2)}{2}.
\end{equation}
It requires a site dependent resolution at the edges of the system but does not involve two-point correlations.
Both quantities were analyzed previously with the aim of identifying the mobility edge in a  Bose-Hubbard model 
in a tilted lattice
\cite{Yao20b}. 
Time evolution of both quantities for $K = 8$, $U_1 = 25$ and different initial state energies are depicted in the
first row of Fig.~\ref{fig:dist_fluct_evolution}. Initial states were prepared in quantum quench of $\phi_0$, 
with $\Delta\phi_0 = 0.50, 1.00, 1.15, 1.35$ and $1.50$, where two first values yield $\epsilon \approx 0.21$
and $0.66$ in the localized regime, third gives $\epsilon \approx 0.78$ in the vicinity of mobility edge, 
fourth gives $\epsilon \approx 0.89$ which is well in extended regime and fifth -- $\epsilon \approx 0.94$,
which is near the high end of spectrum, where the states localize again. Observe that $\Delta x(t)$ 
evolution starts with an abrupt, ballistic-like transient behaviour, but after a time of the order of 
inverse tunneling energy $1/J = 1$ one can observe a slower,  subdiffusive growth. When the Heisenberg time scale 
(which is proportional to the inverse of the mean level spacing) is 
reached, the transport distance value saturates. For energies corresponding to the localized part of the 
spectrum, $\Delta x$ saturates at low values below a single site unit distance while for those in the 
thermal domain $\Delta x$ is of the order of system size $K=8$. Here, in fact, the saturation is reached well 
before the Heisenberg time when the system size is reached. {In the vicinity of the crossover between ergodic and MBL regimes the} saturation of $\Delta x$ 
occurs {only} later and at {an} intermediate value (of the order of unity). Notice, that $\Delta x$ for
the highest
energy considered $0.94$ saturates at lower value then for $0.89$ confirming the presence of the localized regime 
at the high end of the spectrum. We observe a similar situation for the edge fluctuation, $F$. The initial 
transient behavior 
is followed by a typical, logarithmic growth followed by a saturation of $F(t)$.

It was  proposed in \cite{Yao20b} to detect the mobility edge using the notion of pseudovelocity. The authors computed the ratio
\begin{equation}
    V_O = \frac{O(t_\text{max}) - O(t_\text{min})}{ t_\text{max} - t_\text{min}}
\end{equation}
where $O(t) = \Delta x$ or $F$ for fixed times $t_\text{max}$ and $t_\text{min}$
for evolution of states with different energies and plotted them against $\epsilon$. 
They noticed a peak of this value near mobility edge, providing an experimental way for its observation.
However, in our case, the mobility edge is relatively high in the rescaled
energy $\epsilon$ hindering the observation of a dip of pseudovelocity
after the initial rise. Also, for large $K$ and maximal simulation time $t=10^3$ some observables are not yet
saturated.  In effect the pseudovelocities obtained had large fluctuations. In order to mitigate that 
issue, we modified the approach -- 
instead of fixing $t_\text{max}$ and $t_\text{min}$, we scanned $\Delta x(t)$ and $F(t)$ dependencies with $\Delta t = 30$ window 
starting from $t = 10$, computed a linear fit slope in this window and chose its maximal value as a pseudovelocity. 
This allowed us to make $V_O(\epsilon)$ dependencies much smoother (Fig.~\ref{fig:dist_fluct_evolution}c, d). For both $V_x$ 
and $V_F$, on the localized side of spectrum the values are near 0 and start to visibly rise around the 
transition energy $\epsilon \approx 0.75$. They also correctly recognize the shift of transition energy to lower values for 
larger system sizes. Experimental accessibility of $\Delta x$ is lower than $F$, because it requires measuring all the
correlations, while the latter one concerns the sites on the edges only.  {On the other hand $V_F$ is more affected by fluctuations than $V_x$.
Instead of pseudovelocities one may, however, consider simply the average values of observables in a 
fixed time interval as shown in the bottom row of Fig.~\ref{fig:dist_fluct_evolution}. Their behavior is much smoother, providing a similar observation -- a rapid increase of both the mean transport distance and mean fluctuation at the crossover from localized to extended regime.} 

\begin{figure}[htbp]
    \centering
    \includegraphics[width=0.49\linewidth]{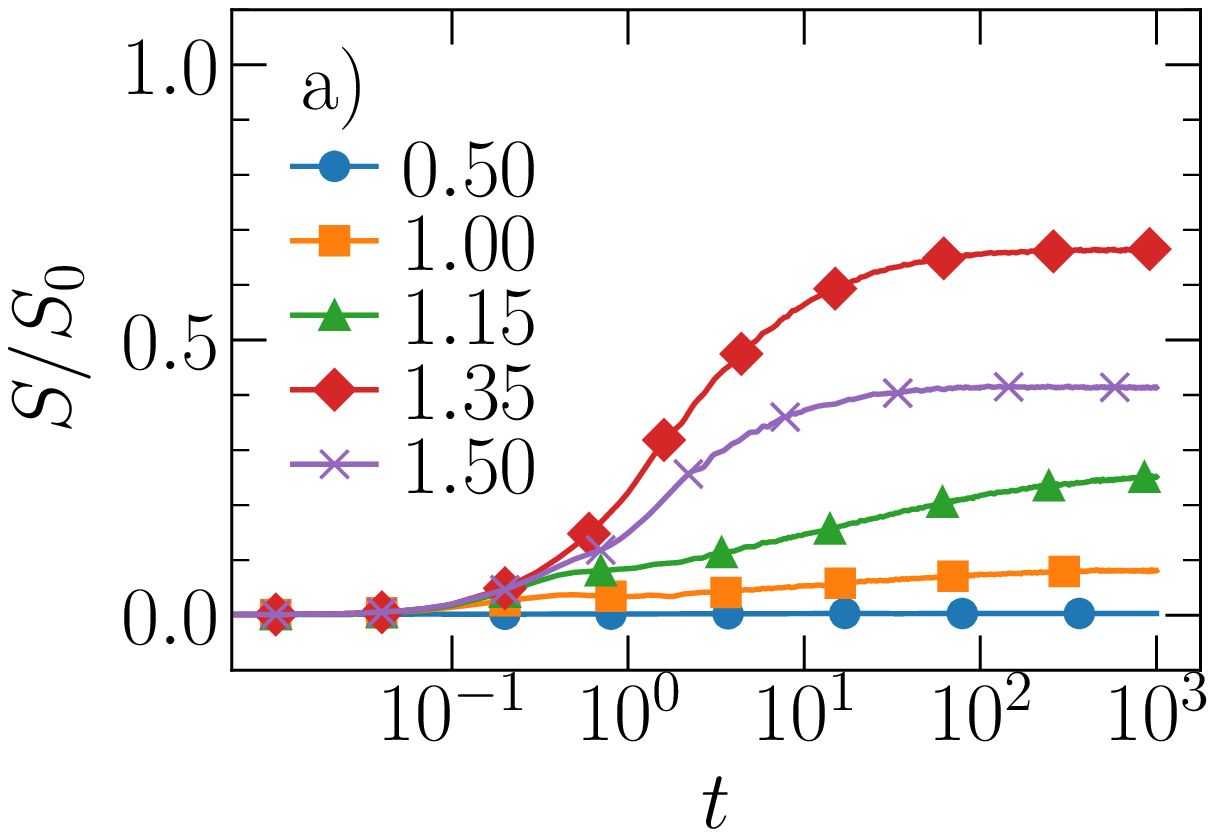}
    \includegraphics[width=0.49\linewidth]{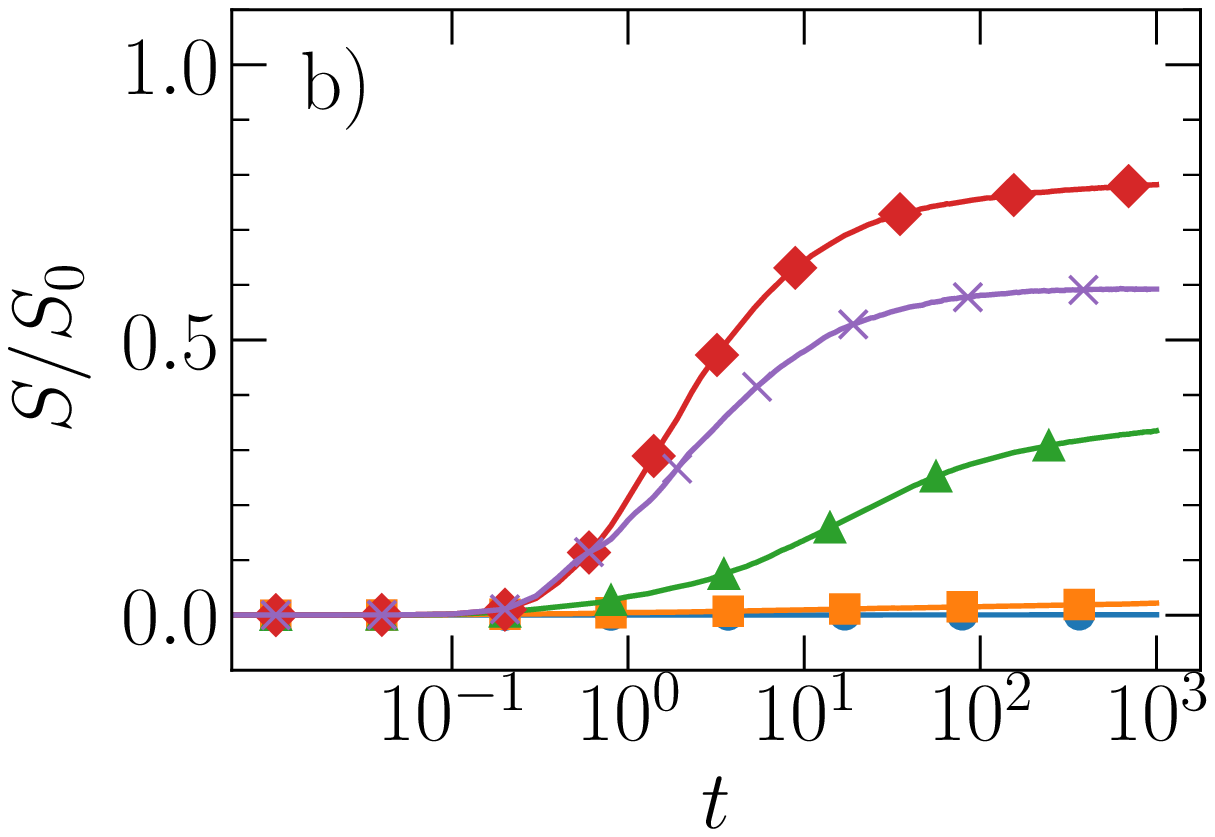}
    \includegraphics[width=0.49\linewidth]{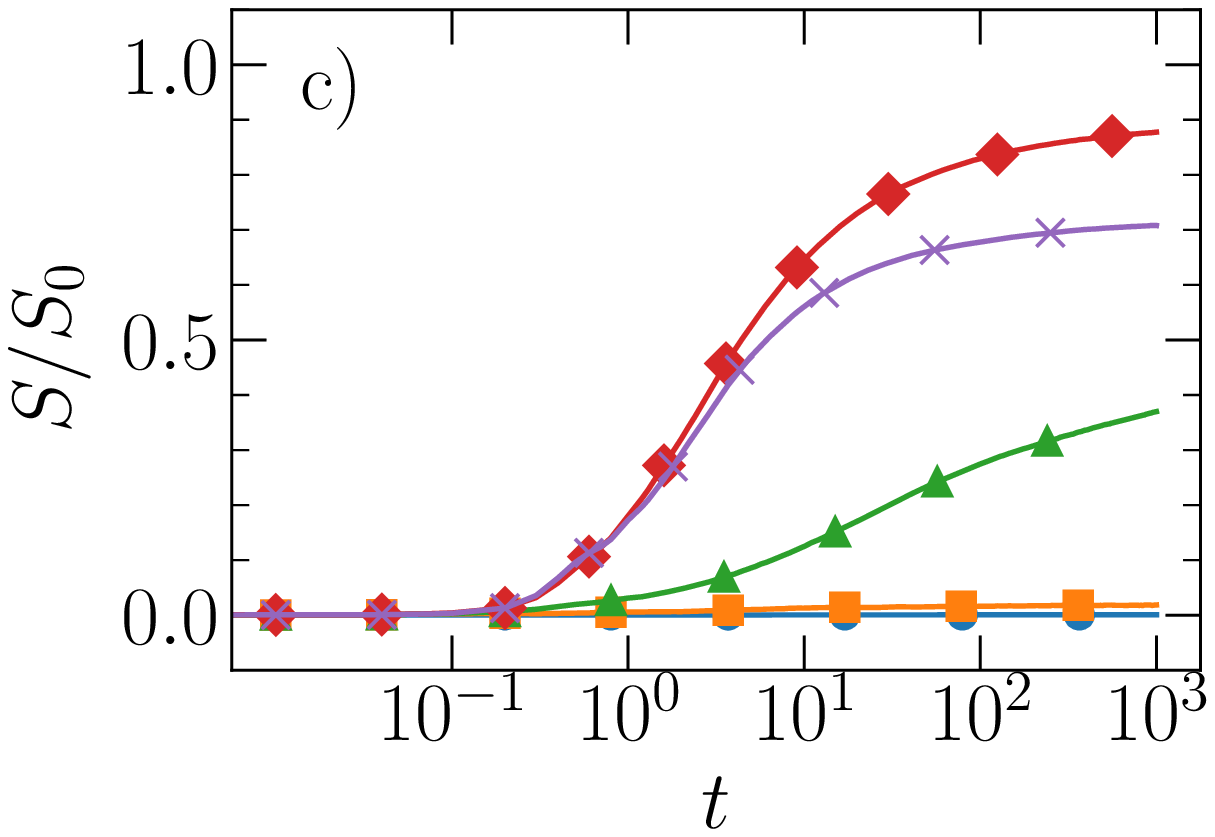}
    \includegraphics[width=0.49\linewidth]{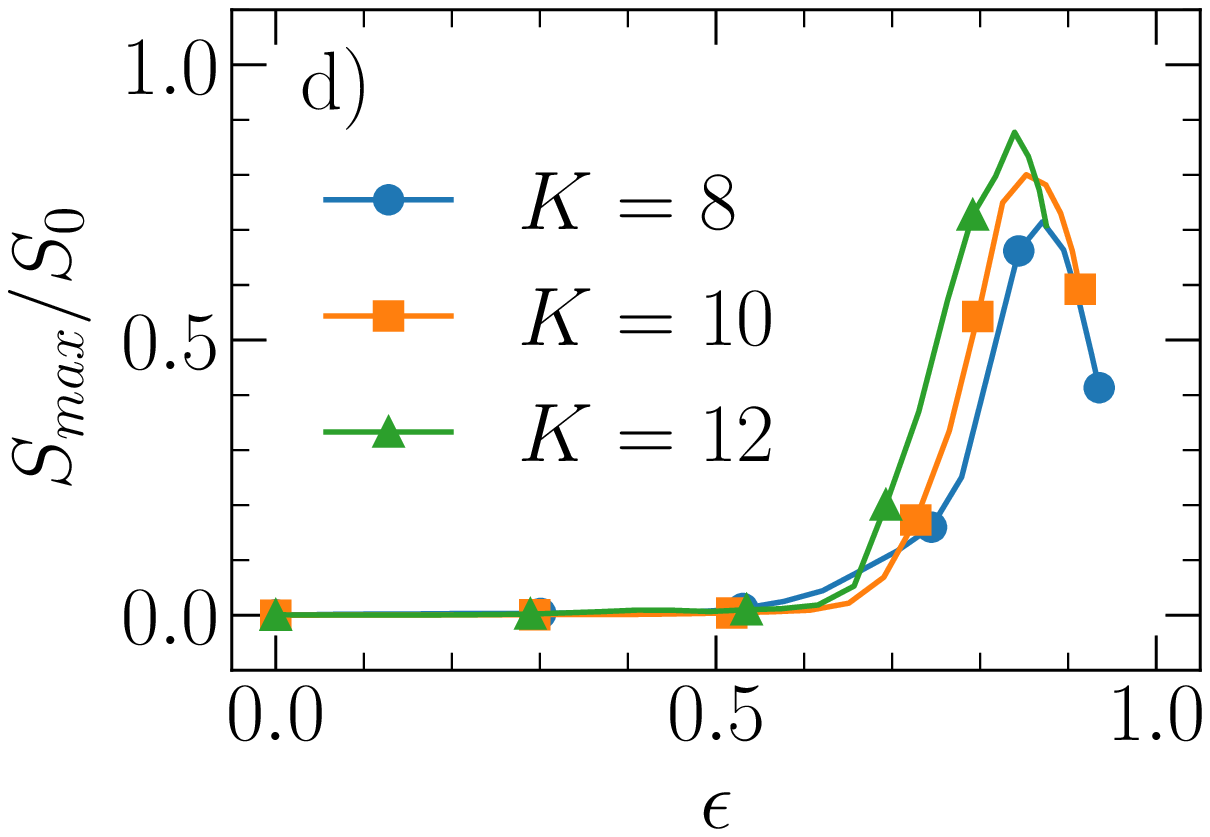}
    \caption{a)-c) Time evolution of von Neumann biparite entanglment entropy for $\Delta\phi_0$-quenched states for, respectively, $K=8,10,12$. d) Maximal value (for $t=10^3$) of $S(t)/S_0$ as a function of energy $\epsilon$ for those 3 system sizes.}
    \label{fig:s_evolution}
\end{figure}

We may also consider time  evolution of the entanglement entropy using the same quenched initial states.
For initially separable state the hallmark of MBL is the logarithmic entanglement entropy growth
\cite{Znidaric08,Bardarson12,Serbyn13a}.
This has been typically tested for separable, product-like initial states. 
It is interesting, therefore, to observe that entropy growth also for our initial entangled states -
see Fig.~\ref{fig:s_evolution}. For localized initial states $S$ grows very slowly.
In the vicinity of the crossover we observe a delayed saturation, while in the extended regime $S$ saturates 
at the value of the order of the average entanglement entropy of a random Gaussian state $S_0$. However,
$S_0$  is never reached for system sizes and times under
consideration. One may notice, that, once again, the saturation value for highest energy $\epsilon = 0.94$ 
is lower that for $\epsilon = 0.89$, which is consistent with previous results. We also plot, in 
Fig.~\ref{fig:s_evolution}(d), the dependence of maximal $S_\text{max}$ against initial state energy. 
In this case, the visible rise of $S_\text{max}$ starts around $\epsilon \approx 0.6$, below the 
transition energy. Nevertheless, it provides additional evidence supporting the mobility edge, at
least for system sizes studied in this paper.

\section{Summary}

We considered signatures of many body localization in the cavity QED model combined with the optical
lattice whose wavelength is incommensurate with the cavity mode wavelength. In such a case the 
effective long-range interactions between atoms (bosons) in the system become dependent on the 
incommensurability ratio resulting in the effective quasi-random potential. Importantly, we do 
not assume any other type of disorder to be present.

We analyse both spectral properties of the system (mean gap ratio as well as entanglement 
entropy of eigenstates) as well as the time dynamics from specially prepared intitial states.
For finite (small, amenable to exact diagonalization) systems we observe clear signatures 
of MBL with mean gap ratio reaching the Poisson level characteristic for a localized regime.
However, the interaction strength leading to localization seems to increase with the system size.
In effect, in the standard thermodynamic limit, we predict that the system will remain delocalized 
for almost all the states. One may define, however, a nonstandard thermodynamic limit, in which 
the interaction strength is not scaled with system size. While this makes the energy of the system a super-extensive
quantity, it also gives rise to  a clear mobility edge
between localized and delocalized regimes.

These findings, based on  spectral statistics, were confirmed and verified by time dynamics in which {we addressed} quantities that are within the reach of current experiments. Interestingly,
meaningful studies required quantum quench spectroscopy with initial states obtained by a 
parameter quench. We have shown that a great care should be taken in the choice of the quenched parameter,
in particular for necessarily small system sizes considered. We have also shown that in the localized
regime the entanglement entropy growth for such initial weakly entangled states follows a logarithmic 
law, a hall mark of MBL observed typically for initial product states.

{ Our results indicate that many-body cavity quantum electrodynamics setups are well suited to studies of quantum ergodicity and its breaking. The setup considered in this work is considerably simpler than the system investigated in \cite{Sierant19c}. Here, the cavity mediated interactions act effectively as a source of randomness leading to a non-trivial interplay between ergodic and MBL eigenstates at different energies. }

\section*{Acknowledgements}

The support by National Science Centre (Poland) under project OPUS 2019/35/B/ST2/00034 (J.Z.) is acknowledged.
P.K. would like to acknowledge the support of Ministry of Science and Higher Education (Poland) grant no. 0108/DIA/2020/49.
P.S. acknowledges the support of  Foundation  for
Polish   Science   (FNP)   through   scholarship   START.
G. M. acknowledges the support by the German Research Foundation (Project-ID 429529648 -- TRR 306 QuCoLiMa "Quantum Cooperativity of Light and Matter) and by theGerman Ministry of Education and Research (BMBF), QuantERA project NAQUAS. Project NAQUAS has received funding from the QuantERA ERA-NET Co-fund in Quantum Technologies implemented within theEuropean Union?s Horizon2020 program.
Numerical simulations were carried out with the support of the Interdisciplinary Center for Mathematical 
and Computational Modeling (ICM) at the University of Warsaw under grant no.\ GB76-1. For linear 
algebra computations, Armadillo C++ language bindings \cite{Sanderson16, Sanderson18} with
Intel\textregistered\ Math Kernel Library as BLAS and LAPACK backend were used.

\appendix

\section{{Details on the Bose-Hubbard model}}

We consider the setup of Fig. \ref{Scheme} in the regime where the cavity mode dynamics is characterized by a faster time scale than the atomic motion. In this limit one can apply a coarse graining to the time evolution \cite{Habibian13,Sierant19c}. Eliminating the cavity degrees of freedom leads to an effective infinite-range interaction term \cite{Schuetz15}. By assuming that the atoms are tightly bound in the lowest band of an optical lattice with wave number $2k_l$, the interaction term can be cast in the form of Eq. \eqref{Hcav}, with \cite{Habibian13,Dogra16,Sierant19c}
\begin{align}
\mathcal Z_\beta(i,\phi_0)=\int \dd{x}\cos(kx+\phi_0) w_i(x)w_j(x)\,,
\end{align}
where $w_i(x)$ is lowest band Wannier function centred at $i^\text{th}$ site. The other quantities are the wave vector of the cavity mode $k$, the interaction strength $U_1$, and an arbitrary phase $\alpha_0$. For a sufficiently deep optical lattice potential, we approximate $w_i(x) \approx  \delta(x-\pi / k_l(i+1/2))$ and obtain:
\begin{equation}
    \mathcal Z_\beta(i,\phi_0)\approx \cos(2\pi\beta i + \phi_0)\,,
\label{Hcav_0}
\end{equation}
with $\beta = k/(2k_l)$. 

This approach assumes that the cavity dynamics adiabatically follows the atomic motion. In particular the coarse-grained cavity field operator is now proportional to the atomic observables
\begin{equation}
\hat a \propto \sum_{i=1}^K\mathcal Z_\beta(i,\phi_0)\hat{n}_i\,.
\end{equation}
This quantity can be revealed by homodyne detection of the field at the cavity output \cite{Black03,Baumann11}. 

\bibliographystyle{apsrev4-1}
%

\end{document}